\documentclass{article}
\usepackage{amsmath}
\usepackage[autostyle=true]{csquotes} 
\usepackage[pdftex]{graphicx}
\usepackage[colorinlistoftodos]{todonotes}
\usepackage[colorlinks=true, allcolors=black]{hyperref}
\usepackage{hyperxmp}
\usepackage{caption} 
\usepackage{sfmath} 
\usepackage{tabularx}
\usepackage[parfill]{parskip}    
\usepackage{setspace} 
\usepackage{enumerate} 
\usepackage{booktabs} 
\usepackage{fancyhdr}
\usepackage{xcolor} 

\usepackage{multirow}
\usepackage{threeparttable}
\usepackage[english]{babel}
\usepackage{amsthm}
\usepackage{float}

\usepackage{graphicx}
\usepackage{subcaption}
\usepackage{mwe}

\usepackage{adjustbox}
\usepackage{subcaption}
\usepackage{geometry}
\usepackage{makecell}
\usepackage{pdflscape}
\usepackage{listings}
\usepackage{amsmath,amsfonts,amssymb,amsthm}
\usepackage[frozencache,cachedir=.]{minted}
\title{Extrapolation of Relative Treatment Effects using Change-point Survival Models}
\author{Philip Cooney, Arthur White }
\date{December 2023}

\begin{document}
\maketitle

\begin{abstract}
\textbf{Introduction} Modelling of relative treatment effects is an important aspect to consider when extrapolating the long-term survival outcomes of treatments. Flexible parametric models offer the ability to accurately model the observed data, however, the extrapolated relative treatment effects and subsequent survival function may lack face validity. \textbf{Methods} We investigate the ability of change-point survival models to estimate changes in the relative treatment effects, specifically treatment delay, loss of treatment effects and converging hazards. These models are implemented using standard Bayesian statistical software and propagate the uncertainty associate with all model parameters including the change-point location. A simulation study was conducted to assess the predictive performance of these models compared with other parametric survival models. Change-point survival models were applied to three datasets, two of which were used in previous health technology assessments. \textbf{Results}  Change-point survival models typically provided improved extrapolated survival predictions, particularly when the changes in relative treatment effects are large. When applied to the real world examples they provided good fit to the observed data while and in some situations produced more clinically plausible extrapolations than those generated by flexible spline models. Change-point models also provided support to a previously implemented modelling approach which was justified by visual inspection only and not goodness of fit to the observed data.  \textbf{Conclusions} We believe change-point survival models offer the ability to flexibly model observed data while also modelling and investigating clinically plausible scenarios with respect to the relative treatment effects.
\end{abstract}

\section*{Introduction}

In health technology assessment (HTA) of treatments for which survival outcomes are different, there is often a requirement to fit survival models which can provide long-term survival predictions, thus enabling calculation of the costs and outcomes across the full time horizon.

Previous authors have described highly flexible parametric models which allow for the modelling of complex hazards functions, however even these models typically do not accommodate specific scenarios which health economic modellers might wish to consider.\cite{Royston.2002,Kearns.2019} Because decision analysts need to investigate the cost-effectiveness of a treatment relative to another treatment (or standard of care), modelling the relative treatment effects over the course of the time horizon is important.\cite{Jackson.2016}
\newpage
A number of scenarios relating to the joint modelling of the comparator and intervention that are of particular interest include:

\begin{itemize}
    \item Treatment Delay (TD) - Hazard function for both treatment and comparator is the same until a certain timepoint
    \item Loss of Treatment Effect (LTE) - Hazard function for treatment and comparator is the same after a certain timepoint
    \item Converging Treatment Effect (CTE) - Hazard ratio converges over time to one (i.e. equal hazards)
\end{itemize}

For certain treatments it could be hypothesized that a delay may be observed after treatment initiation before differences in the survival times of the groups become apparent. In the opposite scenario after a period of initial benefit the treatment effect may no longer be observed and the hazard function for both treatment and comparator are equal. A related scenario is the potential for a measure of treatment effect such as the hazard ratio (HR) to change after a timepoint and converge to 1 in a smooth fashion. Various applications of CTE and LTE in NICE Technology Appraisals are have been documented previously, however, the timepoint after which the change in treatment effect occurs is uncertain often arbitrary chosen. \cite{Kamgar.2022}

In this paper we will describe a class of parametric survival models known as change-point hazard models, whereby the process assumed to generate the hazard function changes at distinct time-points know as change-points. These models offer greater potential flexibility to model changes in relative treatment effects while still allowing for the commonly used and clinically interpretable proportional hazards (PH) assumption.

Change-point survival models have been considered previously as an improvement to the Bagust and Beale approach.\cite{Bagust.2014,Cooney2.2023} The location and number of change-points for a piecewise exponential model were estimated with the final interval providing the extrapolated hazard (which was subject to standard general population mortality to ensure long-term plausibility). In this paper we generalize the class of change-point models to include the more flexible Weibull model and also allow for the inclusion of covariates.

The aim of this paper is to demonstrate how a variety of scenarios relating to changes in treatment effects can be modelled with change-point models. Estimating these models while assuming that the change-point locations are unknown  fully propagates statistical uncertainty, an important requirement for decision analysis. In the Methods section we will describe the notation of the generic change-point survival model and the three scenarios for modelling relative treatment effects that we have highlighted above. We also describe a simulation study which assesses the ability of the models to recover the true parameters and the accuracy in predicting long-term survival data (compared to other parametric survival models). To provide practical applications, we describe three datasets to which we apply specific change-point models. The Results section describes the outcomes of the simulation studies and the data examples. A Discussion section concludes the paper, highlighting the strengths and limitations of the models and further areas of research.

\newpage
\section*{Methods}
\subsection*{General Notation for Change-point Survival Model}
\label{General-Notation}

\subsubsection*{Likelihood of Parametric Change-point Survival Models \label{Like-Parametric-CP}}

A change-point occurs at observation $q$ if $t_1, \dots , t_q$ are generated differently to $t_{q+1}, \dots , t_n$.  Multiple change-points at timepoints can be denoted as a vector $\boldsymbol{\tau}_{1:k}$ (and individual change-points denoted with a single subscript e.g. $\tau_j$), with these $k$ change-points splitting the data into $k + 1$ segments. It should be noted that for these models we do not require that the change-point locations are at discrete points (i.e. event times).

Owing to the potential for covariates, we require that each individual and interval has a specific hazard function $h(t_i,\boldsymbol{\theta}_{ij})$. For each individual we require the cumulative hazard function up until $t_i$. This includes the cumulative hazard for the interval between $t_i$ (assuming it occurs in the $j$th interval) to the previous change-point $\tau_{j-1}$ denoted as $H((t_i - \tau_{j-1}),\theta_{ij})$. The cumulative hazard function for any previous intervals is also required and is denoted as  $H(\tau_g - \tau_{g-1},\boldsymbol{\theta}_{ig})$. The likelihood of the change-point model can be formulated as follows:

\begin{equation*}
L(\boldsymbol{\tau}_{1:k},\boldsymbol{\theta}|\boldsymbol{t}_{1:n}) =  \prod_{i=1}^{n}\Bigg\{ \prod_{j=1}^{k+1} h(t_i,\boldsymbol{\theta}_{ij})^{\delta_{ij}v_i} \exp\bigg\{-\delta_{ij} \bigg[H((t_i - \tau_{j-1}),\boldsymbol{\theta}_{ij}) + \sum_{g=1}^{j-1} H(\tau_g - \tau_{g-1},\boldsymbol{\theta}_{ig}) \bigg] \bigg\}\Bigg\}.
    \label{eq:like-piecewise-exp-bagust}
\end{equation*}

\noindent
with $v_i = 1$ if the $i\text{th}$ subject was observed to have an event and 0 otherwise (censored). If the $i\text{th}$ subject's time (either censored or an event) is within the $j\text{th}$ interval (mathematically $t_i \in (\tau_{j-1}, \tau_{j}]$), then $\delta_{ij} = 1$  and 0 otherwise.

Furthermore $\boldsymbol{\theta}_{ij}$ is a vector of parameters for each individual which in the case of Weibull model are the shape and scale parameters which can have covariates placed upon them. For example let $\boldsymbol{\theta}_{ij} = \{ m_{ij}, a_{i,j} \}$ with $m, a$  scale and shape parameters respectively. To model covariates we introduce matrices $\boldsymbol{\beta}_m= [\boldsymbol{\beta}_{m_{1}} \dots \boldsymbol{\beta}_{m_{k+1}}]$ and $\boldsymbol{\beta}_a= [\boldsymbol{\beta}_{a_{1}} \dots \boldsymbol{\beta}_{a_{k+1}}]$ whose columns are a $p \times 1$ vector each representing the coefficients of one of the $k+1$ intervals. For the $j$th interval $m_{ij} = \exp(\boldsymbol{Z}_{ij}\boldsymbol{\beta}_{m_{j}})$ in which the scale (location) parameter depends on a vector of covariate values $\boldsymbol{Z}_{ij}$ of size $1 \times p$ and and the coefficients of $\boldsymbol{\beta}_{m_{j}}$. The individual shape parameter is calculated as $a_{i,j} =\exp(\boldsymbol{Z}_{ij}\boldsymbol{\beta}_{a_{j}})$.

If we do not specify a treatment effect for the shape parameter we obtain a proportional hazards model, and for each of the $p$ covariates (naturally excluding the intercept) the interval specific hazard ratio for the $q$th covariate is $\text{HR}_\text{jq}= \exp(\beta_\text{jq})$.

\subsubsection*{Data format for Parametric Change-point Survival Models}

To help clarify the notation in Section \ref{Like-Parametric-CP} we will provide illustrative examples of various change-point scenarios with the number of change-points set to 1, which are typically sufficient to fit the observed data. For the purposes of illustration we will assume a dataset with 5 observations, three assigned to treatment and two assigned to a comparator (or baseline) along with ages of each patient (see Code Chunk \ref{code:Simulated-DF}). For the age variable we define another variable called age\_scale, which scales age variable to have a mean of zero and a standard deviation of one. This can improve efficiency when estimating the model. This also simplifies prediction for the population at the mean age as the coefficient for age can be omitted.

\begin{listing}[H]
\begin{minted}[fontsize=\footnotesize,breaklines,framerule=1pt]{R}
 time status trt   age age_scale
 0.08      1   1 77.64      1.14
 0.14      0   0 67.75     -0.10
 1.44      0   1 73.45      0.61
 2.11      1   1 56.36     -1.52
 3.32      0   0 67.52     -0.13
\end{minted}
\caption{Example Simulated Dataset}
\label{code:Simulated-DF}
\end{listing}

If we split the assuming a change-point at time 1 we obtain multiple subrecords at each change-point. The new dataset are be in ``counting process'' format, with a start time (tstart), stop time (time), and event (status) for each record (Code Chunk \ref{code:Counting-DF}). If the individual survives past 1 year they will have two subrecords the first subrecord will have a tstart = 0, time = 1 and status = 0. The second subrecord will have tstart = 1, time equal to the time in the original dataset and status equal the status as per the original dataset. A column called id indicates which subrecords belong to which patient. Additionally we have added an Intercept column consisting of 1's so that the Intercept, trt and age\_scale columns represent a design matrix for the model parameters. It can be seen how each row of this design matrix represents $\boldsymbol{Z}_{ij}$.

\begin{listing}[H]
\begin{minted}[fontsize=\footnotesize,breaklines,framerule=1pt]{R}
  tstart time status id Interval Intercept trt age_scale
       0 0.08      1  1        1         1   1      1.14
       0 0.14      0  2        1         1   0     -0.10
       0 1.00      0  3        1         1   1      0.61
       1 1.44      0  3        2         1   1      0.61
       0 1.00      0  4        1         1   1     -1.52
       1 2.11      1  4        2         1   1     -1.52
       0 1.00      0  5        1         1   0     -0.13
       1 3.32      0  5        2         1   0     -0.13
\end{minted}
\caption{Example Dataset in a counting process format}
\label{code:Counting-DF}
\end{listing}

\subsubsection*{Change-point models with discrete change in HR}

By restricting various covariates (i.e. elements of $\boldsymbol{\beta}$) to be equal to 0 or equal across the intervals we can specify many different change-point models.

In Code Chunk \ref{code:Beta-covar} we have specified covariates for the scale parameter based on treatment status and age. The effect of age is constant across the intervals while the treatment effect varies across intervals, in fact because it is zero in the second interval the hazards are equal to the comparator i.e. LTE model. The opposite scenario is where we constrain the treatment coefficient to be zero in the first interval yielding a TD model.

The shape parameter is constant across the intervals (and more generally not subject to a treatment effect) resulting in a proportional hazards change-point model. Because the intercept for the scale parameter is constant across intervals, the baseline hazard is continuous at the change-point, however, if the intercept is allowed to vary across intervals, it is non-continuous but still a proportional hazard model.

\begin{listing}
\begin{minted}[fontsize=\footnotesize,breaklines,framerule=1pt]{R}
beta_scale
          Interval-1 Interval-2
Intercept       -0.5       -0.5
trt             -0.2        0.0
age_scale        0.1        0.1

beta_shape
          Interval-1 Interval-2
Intercept       -0.4       -0.4
trt              0.0        0.0
age_scale        0.0        0.0

 tstart time status id Interval Intercept trt age_scale scale shape
       0 0.08      1  1        1         1   1      1.14  0.56  0.67
       0 0.14      0  2        1         1   0     -0.10  0.60  0.67
       0 1.00      0  3        1         1   1      0.61  0.53  0.67
       1 1.44      0  3        2         1   1      0.61  0.64  0.67
       0 1.00      0  4        1         1   1     -1.52  0.43  0.67
       1 2.11      1  4        2         1   1     -1.52  0.52  0.67
       0 1.00      0  5        1         1   0     -0.13  0.60  0.67
       1 3.32      0  5        2         1   0     -0.13  0.60  0.67
\end{minted}
\caption{$\boldsymbol{\beta}$ covariate matrices for shape and scale parameters and updated dataset}
\label{code:Beta-covar}
\end{listing}

\begin{figure}[H]
    \centering
    \includegraphics{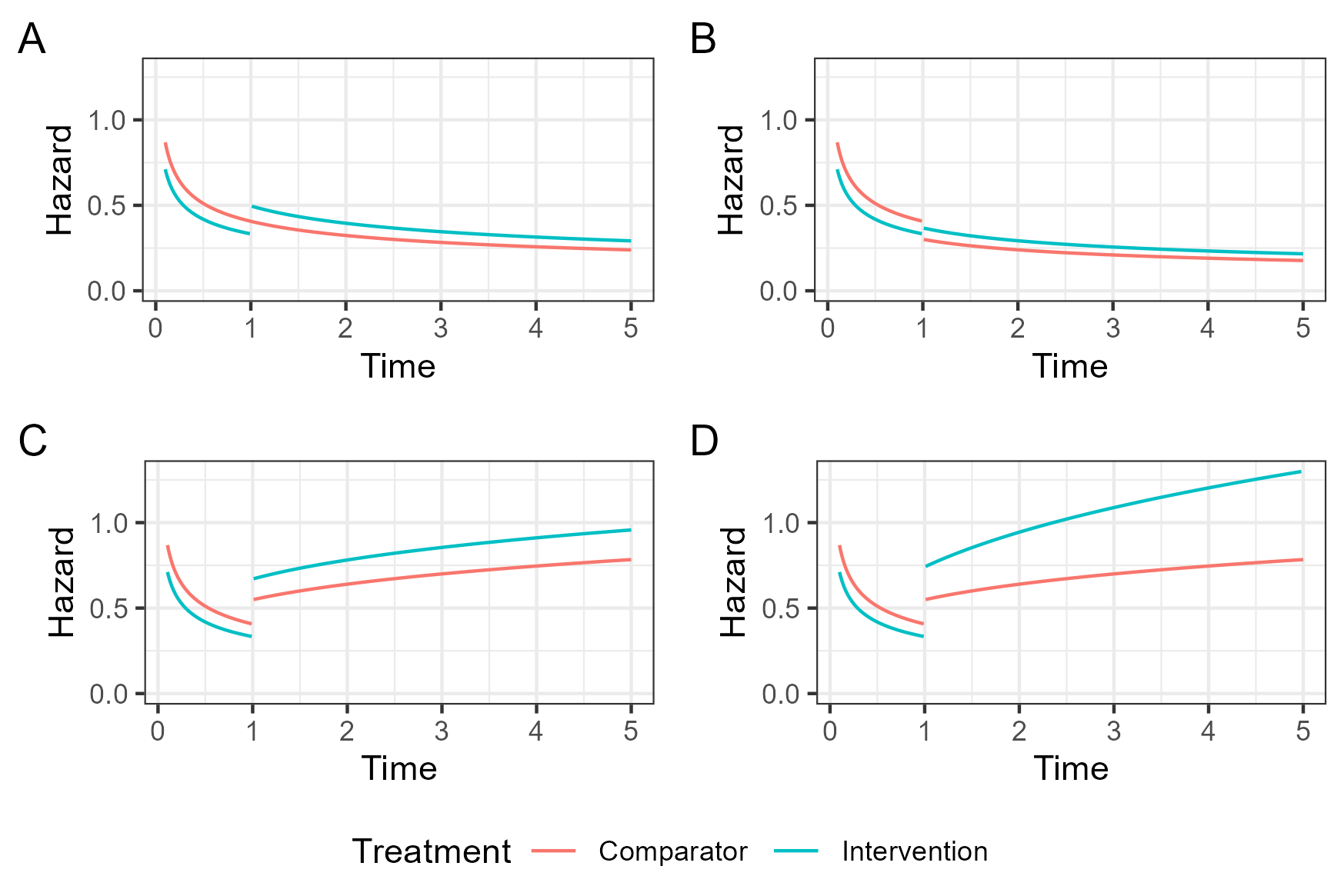}
    \caption{Various scenarios for modelling the hazard function with a change-point}
    \label{fig:HR-step}
\end{figure}

In Figure \ref{fig:HR-step} we present four possible scenarios to jointly model the hazard function for a change-point model in which the HR for the treatment was $< 1$ up until (and including) the change-point and greater than 1 after the change-point. Figure \ref{fig:HR-step}-A illustrates the scenario previously described in which only the HR of the treatment changes after the change-point and so the baseline (comparator) hazard function is continuous. In  Figure \ref{fig:HR-step}-B the intercept also changes across intervals so that there is a different baseline hazard function for each interval. Figure \ref{fig:HR-step}-C extends \ref{fig:HR-step}-B so that the shape parameter also changes across intervals. It is worth highlighting that scenarios A-C still assume proportional hazards while the final scenario in Figure \ref{fig:HR-step}-D assumes different shapes for both treatment arms after the change-point and therefore is no longer a PH model for the interval after the change-point.

\subsubsection*{Change-point scenarios with Convergence of the Hazard Ratio}

In the previous scenario we have considered a step change in the HR, however, we may also allow the HR of the treatment arm to converge in a continuous manner to the comparator or baseline hazard i.e. CTE models. For a converging hazards model we consider a change-point $\tau_{\text{wane}}$ after which the hazard ratio for the treatment from the previous interval ($\text{HR}_{\text{initial}}$) begins to wane (i.e. converge to 1 over time). The HR for any time after $\tau_{\text{wane}}$ is
\begin{equation}
   \text{HR}(t) = 1 - (1- \text{HR}_{\text{initial}})\exp(-\omega (t-\tau_{\text{wane}} )),
   \label{eq:converge-HR}
\end{equation}
 were $\omega$ is a constant rate at which the HR converges to 1.  

In order to estimate the cumulative hazard function for the treatment arm, we need to evaluate $\int_{\tau_{\text{wane}}}^t \text{HR(t)}\text{h}_{\text{baseline}}(t) dt$ with $\text{h}_{\text{baseline}}(t)$ being the baseline hazard. For the Weibull model the indefinite integral is  $am_{\text{baseline}}(t^{a})\Big(1/a-(\Gamma(a,\omega t)(\text{HR}_{\text{initial}}-1)\exp(\omega\tau_{\text{wane}}))/(\omega t)^{a}\Big) + C$  for the interval beyond $\tau_{\text{wane}}$.  $m_{\text{baseline}}$ refers to the baseline scale parameter with the shape $a$ common for both intervention and comparator arm and $\Gamma(a,\omega t)$ is a the upper incomplete gamma function. For the exponential likelihood the integral simplifies considerably with $a = 1$.

We define the $\boldsymbol{\beta}_m$ matrix as before see Code Chunk \ref{code:Beta-covar-CTE}, however, the HR for the treatment effect beyond the change-point is a function of time since the change-point, the HR for the treatment before the change-point and  $\omega$, the rate of convergence (Equation \ref{eq:converge-HR}).

\begin{listing}[H]
\begin{minted}[fontsize=\footnotesize,breaklines,framerule=1pt]{R}
beta_scale
          Interval-1 Interval-2
Intercept       -0.5       -0.5
trt             -0.2        log(HR(t))
age_scale        0.1        0.1
\end{minted}
\caption{$\boldsymbol{\beta}_m$ covariate matrix for CTE model}
\label{code:Beta-covar-CTE}
\end{listing}

\begin{figure}[H]
    \centering
    \includegraphics{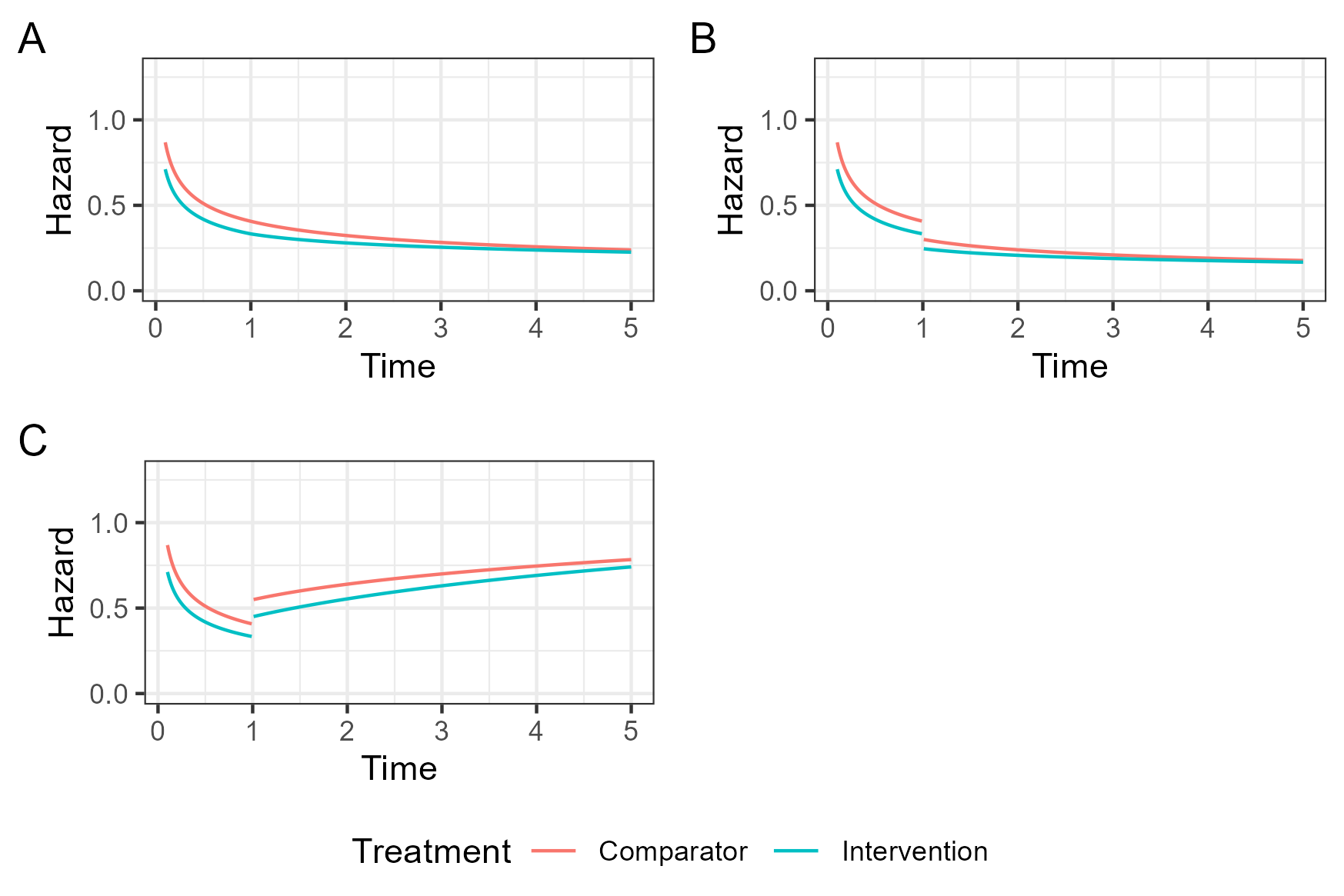}
    \caption{Various scenarios for converging hazard with a change-point}
    \label{fig:HR-wane}
\end{figure}

Each of the scenarios presented in Figure \ref{fig:HR-wane} correspond to those (proportional hazard models) presented in Figure \ref{fig:HR-step}.

\subsection*{Estimation of Change-point models}

All of the models described in the subsequent sections are estimated using the JAGS statistical software programme.\cite{Plummer.2003} The model code to define the change-point model likelihood makes extensive use of the $\texttt{step(x)}$ function, which is a function which returns a value of 1 if x $\geq 0$ and  0 otherwise. Using the notation defined previously we can use two step functions to calculate the interval which $t_i$ is in. To test if $t_i$ is within the $k$th interval we can use the following $\text{step}(t_i - \tau_{k-1} -c)*\text{step}(\tau_{k}-t_i)$ which will return 1 if and only if $t_i$ is in the $k$th interval. Technically the first step function requires $-c$ where $c$ is a very small number as $t_i$ should be strictly greater than $\tau_{k-1}$ to be in the $k$th interval. In practical terms this factor can be excluded when dealing with continuous $\tau$ as the probability $\tau = t_i$ is 0. This allows for the calculation of the $\delta_{ij}$ function required to identify the change-point interval. Because JAGS does not have distributions corresponding to a Weibull change-point model we need an approach to include the likelihood contribution of each datapoint. In order to define such a contribution we use the ``zeros trick''. The zeros trick is to assume a Poisson (phi) observation of zero has likelihood exp(-phi), so if our observed data is a set of 0's, and phi[i] is set to -log(L[i]), we will obtain the correct likelihood contribution. Note that phi[i] should always be $>$ 0 as it is a Poisson mean, and so we may need to add a suitable constant to ensure that it is positive.

For the change-point we assume that $\tau_{1:k}$ are even ordered statistics of 2$k$ split points drawn from a Uniform distribution on $[0, \tau_{\text{max}}]$ and we assume that  $\tau_{\text{max}}$ is the maximum observed time. Formally this prior distribution is $p(\boldsymbol{\tau}_{1:k}|k)= \frac{(2k + 1)!\prod_{k = 1}^{k +1}(\tau_k - \tau_{k-1})}{\tau_{\text{max}}^{2k+1}}$ and has been used extensively in the estimation of  continuous change-point models.\cite{Chapple.2020}

It also possible to estimate change-point models assuming when assuming that time is partitioned into a number of intervals. By calculating the number of deaths and at-risk population time in each interval a Weibull model can be estimated as a generalized linear model with a Poisson likelihood.\cite{Kearns.2019} Although the values that the change-points can take are now discrete and all covariates in the model are required to be categorical model estimation is much quicker, particularly for large sample sizes. Other models such as Gompertz, Log-logistic and Log-Normal can also be estimated in the generalized linear model framework.

Codes to reproduce the analysis presented in this paper are available on \href{https://github.com/Anon19820/Flex-Changepoint/tree/main}{\textcolor{blue}{\underline{$\texttt{Github}$}}} with skeleton pseudo-code provided (Code Chunk \ref{code:JAGS-cp-code}) for the special case of a one change-point Weibull model with no covariates in the Appendix.

\subsection*{Simulation Study}

We consider three change-points models for generating data arising for change-point models under a variety of different parameters and sample sizes. In all scenarios we assume that the baseline scale parameter for the Weibull model is 0.3. The change-point itself is assumed to be one of either  $\tau = \{1,2\}$. The model assumes common shape parameter for both intervals being either 0.7 or 1.2 (assuming monotonically increasing or decreasing hazards). We assume the HR between the treatment and baseline is 0.25, 0.5 or 0.75. The sample size considered for each arm was assumed to be 100, 300 or 500 ($n_\text{samp}$) and the data-cut off was either 3 or 5 years ($t_{\text{cens}})$ after which all observations were assumed censored, with no censoring before the data-cut off.

In the first set of scenarios we assume a treatment delay. Before a change-point the hazards for both treatment and baseline are equal. After the change-point we assume a PH model with various HRs and baseline hazard functions investigated. The scenarios in which a monotonically decreasing and increasing baseline hazard are assumed are considered in Figures \ref{fig:Scenario-TD}.

In the second set of scenarios we assume a loss of treatment effect after the change-point which occurs at 2 years, while before the change-point the data arise from a PH model. The scenarios in which a monotonically decreasing and increasing baseline hazard are assumed are considered in Figures \ref{fig:Scenario2-LTE}.

For each of the combination of parameters within the scenarios the models were estimated on 50 simulated datasets. Across each of the 100 simulated datasets we are interested in two quantities. Firstly we are interested in the expected difference in the survival between both arms across the relevant time horizon ($t_{max}$) which we assumed was 15 `` years '' and i.e. the difference in Restricted Mean Survival Time for both arms denoted $\text{RMST}_{\text{diff}}$. We subtract the model's $\text{RMST}_{\text{diff}}$ from the true $\text{RMST}_{\text{diff}}$, and the absolute value of this value is then averaged over the 100 simulated datasets and denoted as $\text{Err}_{\text{diff}}$. Values of $\text{Err}_{\text{diff}}$ which are closer to zero indicate that a particular model was on average closer to the true $\text{RMST}_{\text{diff}}$.  We also assessed $\text{Err}_{\text{diff}}$ for a range of standard parametric models i.e. Exponential, Weibull, Gamma, Gompertz, Log-Logistic, Log-Normal and Royston-Parmar models. For the Royston-Parmar model we considered two specifications a PH model and a non PH model. For the non-PH Royston-Parmar model we placed a treatment effect on the $\gamma_1$ parameter allowing for comparable flexibility to the change-point model.

Secondly we are interested in how the model can recover the true parameters. To do this we calculated the posterior median of the parameters. Across the 100 datasets we calculated the mean and standard deviation of these posterior medians. This expected value shows how close the posterior medians are to the true parameter, while the standard deviation provides a measure of standard error for the parameter.

Because the simulation study required us to run a large number of models we set the number of iterations for each of the two chains to each with 20,000 with after a burn-in of 2000 iterations. Every fourth iteration was retained (i.e. thinning of 4) to give a total posterior sample of 10,000 iterations across the two chains.

\subsection*{Practical Applications to Clinical Trial Data}
\label{Example-Dataset}

In this section we provide some background on the datasets used and the hypotheses to be tested. Regarding the For each of the change-point models described in this section we ran 2 chains for 55,000 iterations with an initial burnin of 5,000 and a thinning factor of 5 (thus giving us a total sample of 20,000 iterations).  Convergence diagnostics were assessed using the $\texttt{ggmcmc}$ package.\cite{ggmcmc.2016}

\subsubsection*{E1690 \& E1684 - Multiple Change-point Scenarios}

An immunotherapy known as interferon $\alpha$-2b was evaluated in two observation-controlled Eastern Cooperative Oncology Group (ECOG) phase III clinical trials, E1684 and E1690. The first trial, E1684, was a clinical trial comparing high-dose interferon (IFN) to Observation (OBS). A further confirmatory study, E1690 was initiated in 1991 to attempt to confirm the results of E1684. Various analyses of these trials are presented elsewhere.\cite{Ibrahim.2001}

By combining the E1684 and E1690 (as was also considered previously),\cite{Ibrahim.2001} we obtain a dataset with long term survival data of a group of patients treated with immunotherapy for multiple myeloma (up to 10 years). This long term dataset allows us to consider various scenarios, including that the relative treatment effect dissipates, possibly because patients are no-longer receiving treatment. Although not considered in a technology appraisal, this dataset has the previously stated advantage of having a very long term follow up along with information on potential other covariates of interest which are not available when we digitize published Kaplan-Meier from technology appraisals.

Of interest in a change-point analysis, there is evidence of violation of the proportional hazard assumption for the treatment, but not for other covariates such as age as assessed by Schonefeld residuals. As noted previously, change-point models can investigate a variety of scenarios with respect to the hazard ratio for the treatment effect while also including covariates which do satisfy the proportional hazard assumption.

In the first scenario we consider a Weibull model with a change-point in the hazard ratio for treatment (INF) vs control (OBS) - Scenario 1. A second scenario considers LTE Weibull change-point model, noting that this differs from the first scenario in that the HR for the second interval is constrained to be equal to 1 - Scenario 2. We then consider CTE model - Scenario 3. The relative goodness of fit of each of these survival models will be presented.

Finally, we will compare estimates from the parametric change-point model with those estimated using a Cox (semi-parametric) change-point model.

For these scenarios we assume that there is a different baseline hazard function for each interval (i.e. scenario presented in Figure \ref{fig:HR-step}-B or C) rather than a common baseline hazard function (i.e. scenario presented in Figure \ref{fig:HR-step}-A).  For each of datasets considered we also fit standard parametric models, calculating the statistical goodness of fit and the $\text{RMST}_{\text{diff}}$ for all models (also conducted for the other examples). Assessment of relative goodness of fit to the observed data was estimated using WAIC.

\subsubsection*{LUME-LUNG 1  - Potential Treatment Delay}

In \cite{RN21} the technology of interest was nintedanib + docetaxel vs docetaxel monotherapy in the adenocarcinoma population. Within this population the Kaplan-Meier curves for overall survival do not appear diverge until $\approx$ 5 months. \cite{Reck.2014} If we fit a standard parametric Weibull (PH) model to the data, the estimated survival assuming proportional hazards may not accurately model the initial section of the data in which  the survival curves are very similar. However, a change-point model in which a common Weibull model followed by a Weibull model allowing for a different HR with respect to treatment could allow for a better fit to the data and potentially a more plausible extrapolation. As for the E1690 \& E1684 dataset we assumed the change-point model had different hazard functions for both of the intervals.

\subsubsection*{BRIM-3 Study - Loss of Treatment effect}

\begin{figure}[H]
\centering
\includegraphics[width=0.9\textwidth]{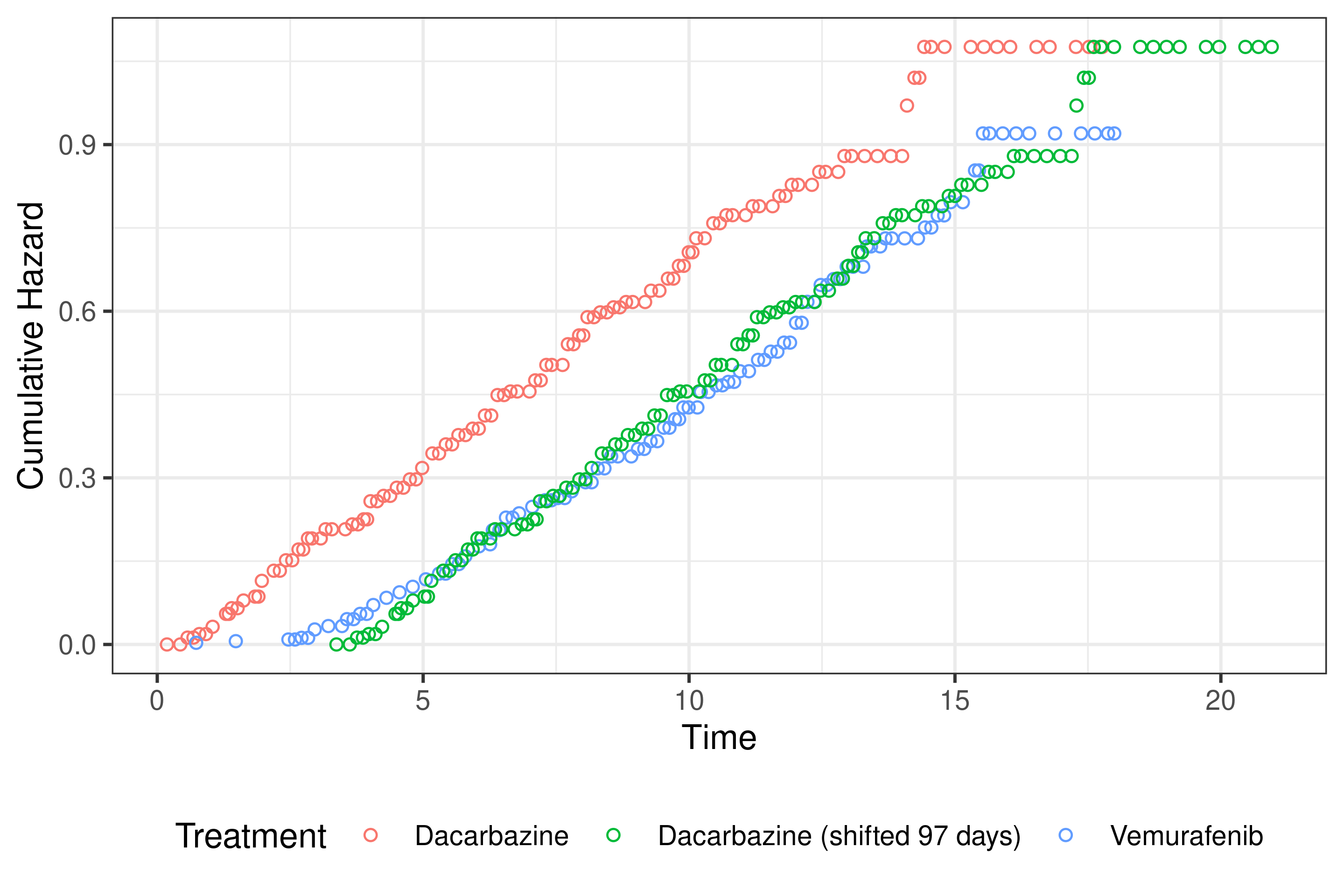}
\caption{BRIM-3 Trial OS cumulative hazard plot.}
\label{fig:TA269-Bagust-Beale}
\end{figure}

It has been suggested that the effect of vermurafenib in the BRIM-3 trial is restricted to the first three months of the clinical trial after which a constant common hazard is apparent.\cite{Bagust.2014,BRIM-3} They conclude this by inspecting the cumulative hazard plot shown in Figure \ref{fig:TA269-Bagust-Beale} in which they shift the cumulative hazard of the dacarbazine arm 3 months and note that it approximately lines up with the cumulative hazard of vermurafenib.

It may be of interest to consider whether a constant hazard model or Weibull model fits the data best. We could assess this by fitting two change-point models to the data, one with a constant hazard and another using a Weibull model. In these models, the change-point will apply only to the vemurafenib arm, with the hazard after the change-point estimated from the data beyond the change-point for the vemurafenib arm and the entire data for the dacarbazine arm.

In terms of the parameters of a change-point model this model can be estimated by allowing a common intercept across intervals for both the shape and scale (See parameters $x,z$ in Code Chunk \ref{code:Beta-covar-TA269}. For interval 1 both the shape and scale are subject to a treatment effect (denoted by parameters $y,w$), while for interval 2 there is none, denoted by 0. This extends the previous LTE models in that we have a non-proportional hazard model for the first interval then equal hazards after the change-point along with a continuous function for the baseline hazard.

\begin{listing}
\begin{minted} [fontsize=\footnotesize,breaklines,framerule=1pt]{R}
beta_scale
          Interval-1 Interval-2
Intercept       x       x
trt             y       0.0

beta_shape
          Interval-1 Interval-2
Intercept       z       z
trt             w       0.0
\end{minted}
\caption{$\boldsymbol{\beta}$ covariate matrices for shape and scale parameters in loss of Treatment Effect}
\label{code:Beta-covar-TA269}
\end{listing}

\newpage
\section*{Results}

\subsection*{Simulation Study Results}

We present a subset of the simulation study results in Appendix \ref{Appendix:Sim-Res} with full results available in the Github Repository. A number of points are worth noting. For each of the scenarios we observe that the $\text{Err}_{\text{diff}}$ is lowest for the change-point models with lowest with large sample sizes and smaller values of the HR. A HR of 0.75 produced $\text{Err}_{\text{diff}}$ comparable to the next best models, suggesting a relatively large HR is required to adequately estimate the model.

The results are also sensitive to the baseline hazard. This is because for the scenarios which have a monotonically decreasing hazards result in a higher proportion of the sample being censored at the end of follow up. As a greater proportion of the sample needs to be extrapolated the error in $\text{Err}_{\text{diff}}$ is also greater.

Across the range of parameters considered for the converging hazards model the $\text{Err}_{\text{diff}}$ was only moderately lower than the next best model. Additionally the standard error for the parameters were quite large relative to the other scenarios.

\subsection*{E1690 \& E1684 - Various Change-point Hypotheses}

For all analysis detailed in this subsection, a covariate for age is included whose HR is fixed with respect to time. In order to plot the survival function stratified by treatment we require the age variable to be set at a particular value (i.e. the survival functions vary with respect to age). In all results presented, hereafter, the survival stratified by treatment is predicted at the mean value of the age from the combined trial  (i.e. setting age\_scale = 0).

In each of the figures presenting the survival functions for the change-point models  (including those in the subsequent subsections) the posterior density of the change-point was presented (green density with red outline).

\subsubsection*{Scenario 1 - Weibull Model step change in HR}
\label{scenario-1}
In the first scenario we consider a step change in the HR, with the baseline shape and scale parameters also changing before and after the change-point (Figure \ref{fig:Scenario1}).

The median change-point is 1.19 years (with a 95\% credible interval 0.73-1.90)  after which the median HR is 1.07 (with a 95\% credible interval 0.8-1.44), suggesting that the treatment effect dissipates (median HR before the change-point was 0.63 with 95\% credible interval 0.33-0.93).

\subsubsection*{Scenario 2 - Equal hazards after change-point}

In the second scenario we assume that before the change-point there we have a proportional hazards Weibull model and after the change-point the hazards are generated from a common Weibull model with a different baseline shape and scale (Figure \ref{fig:Scenario2}).  In this scenario the mean change-point was 1.2 years.

\subsubsection*{Scenario 3 - Converging hazards after change-point}

A third scenario was the assumption of a converging hazard in which the hazard ratio between the treatment and the intervention converges to a value of 1. The average time of the change-point after which the hazard begins to converge is 1.1 years and the HR before the change-point of had a median value of  0.55 (with 95\% credible interval 0.25-0.90). 

\begin{figure}[H]
        \centering
        \begin{subfigure}[b]{0.475\textwidth}
            \centering
            \includegraphics[width=\textwidth]{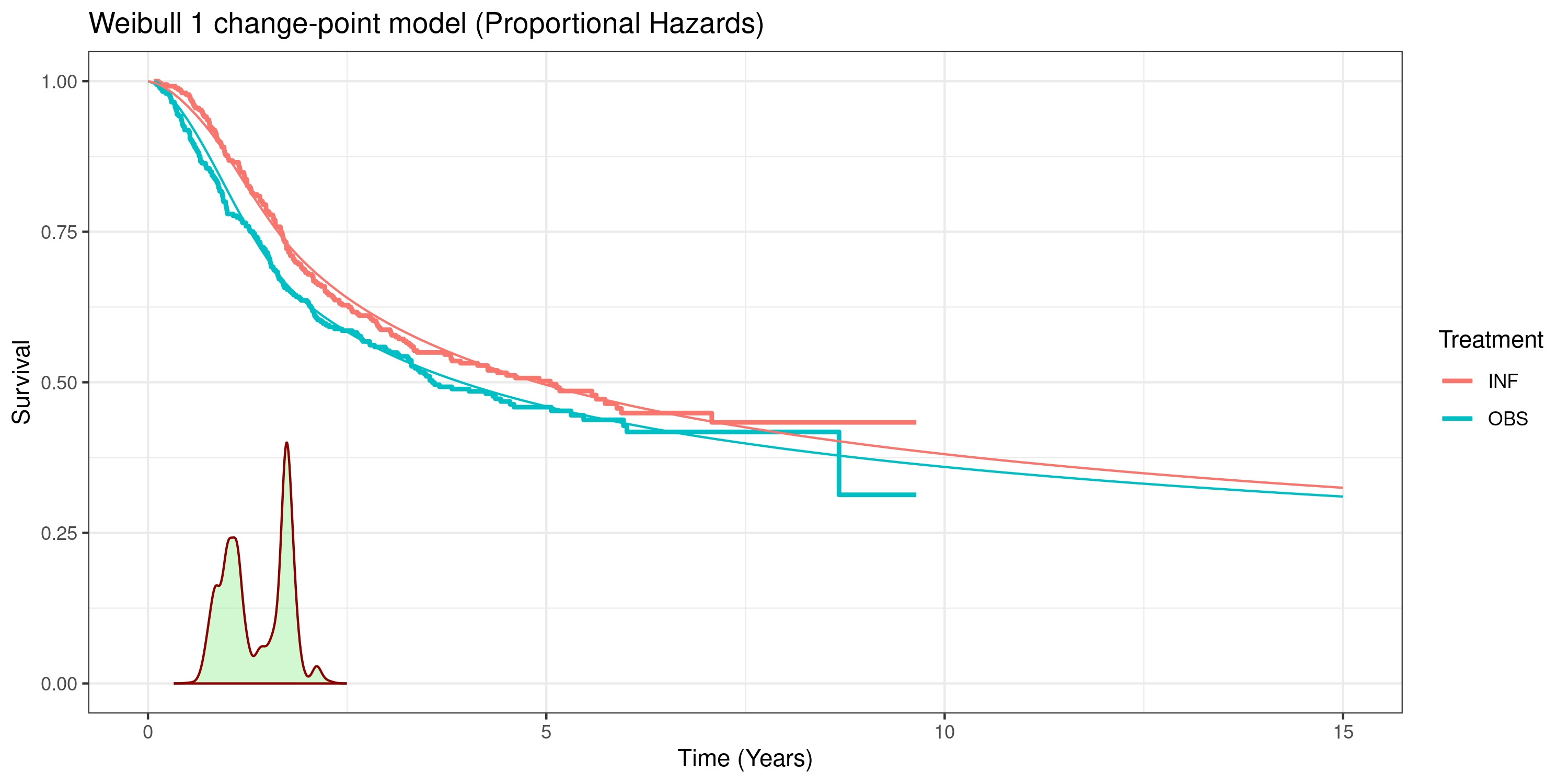}
            \caption{Scenario 1}%
            \label{fig:Scenario1}
        \end{subfigure}
        \hfill
        \begin{subfigure}[b]{0.475\textwidth}
            \centering
            \includegraphics[width=\textwidth]{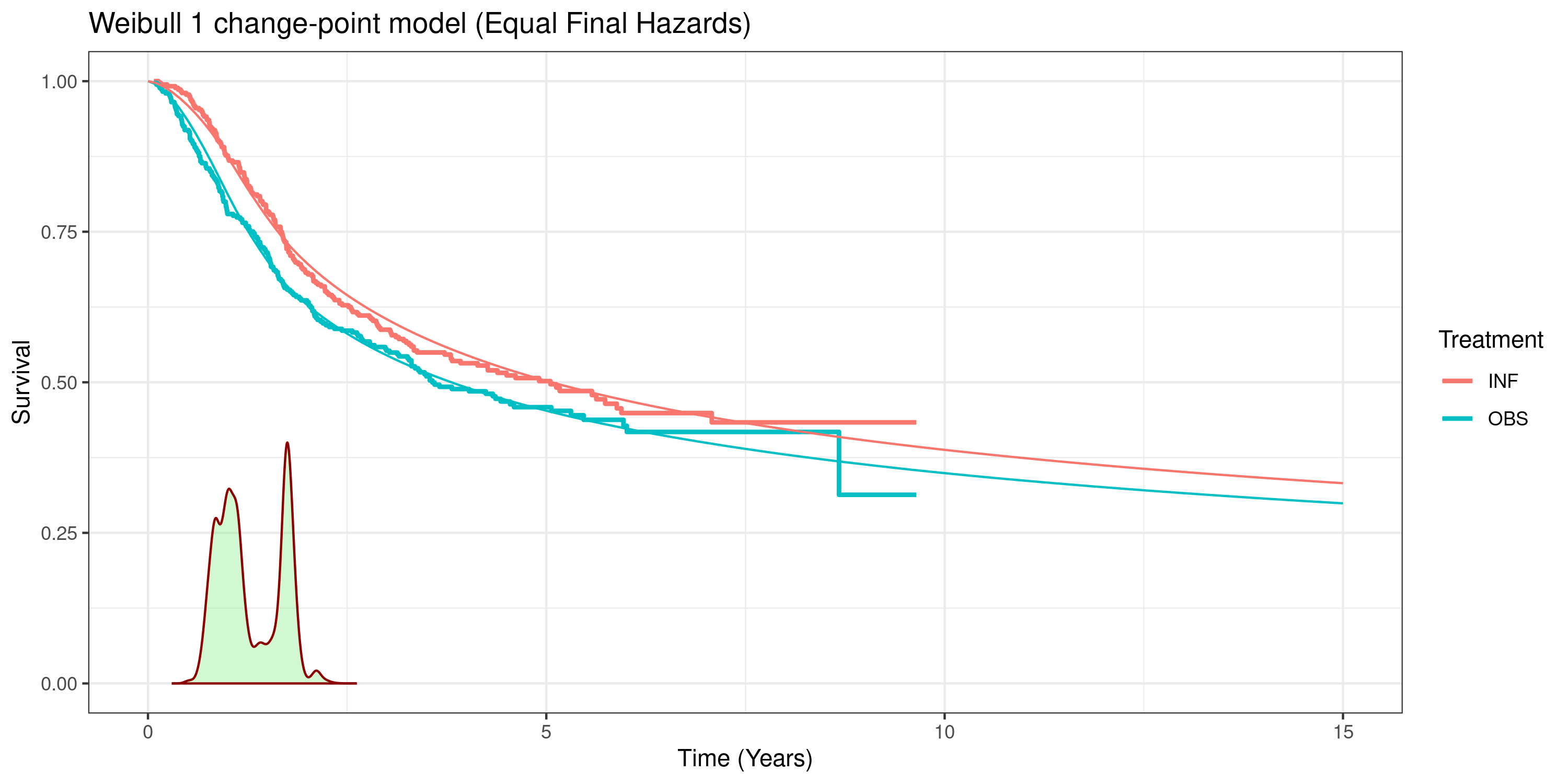}
            \caption{Scenario 2}%
            \label{fig:Scenario2}
        \end{subfigure}
        \vskip\baselineskip
        \begin{subfigure}[b]{0.475\textwidth}
            \centering
            \includegraphics[width=\textwidth]{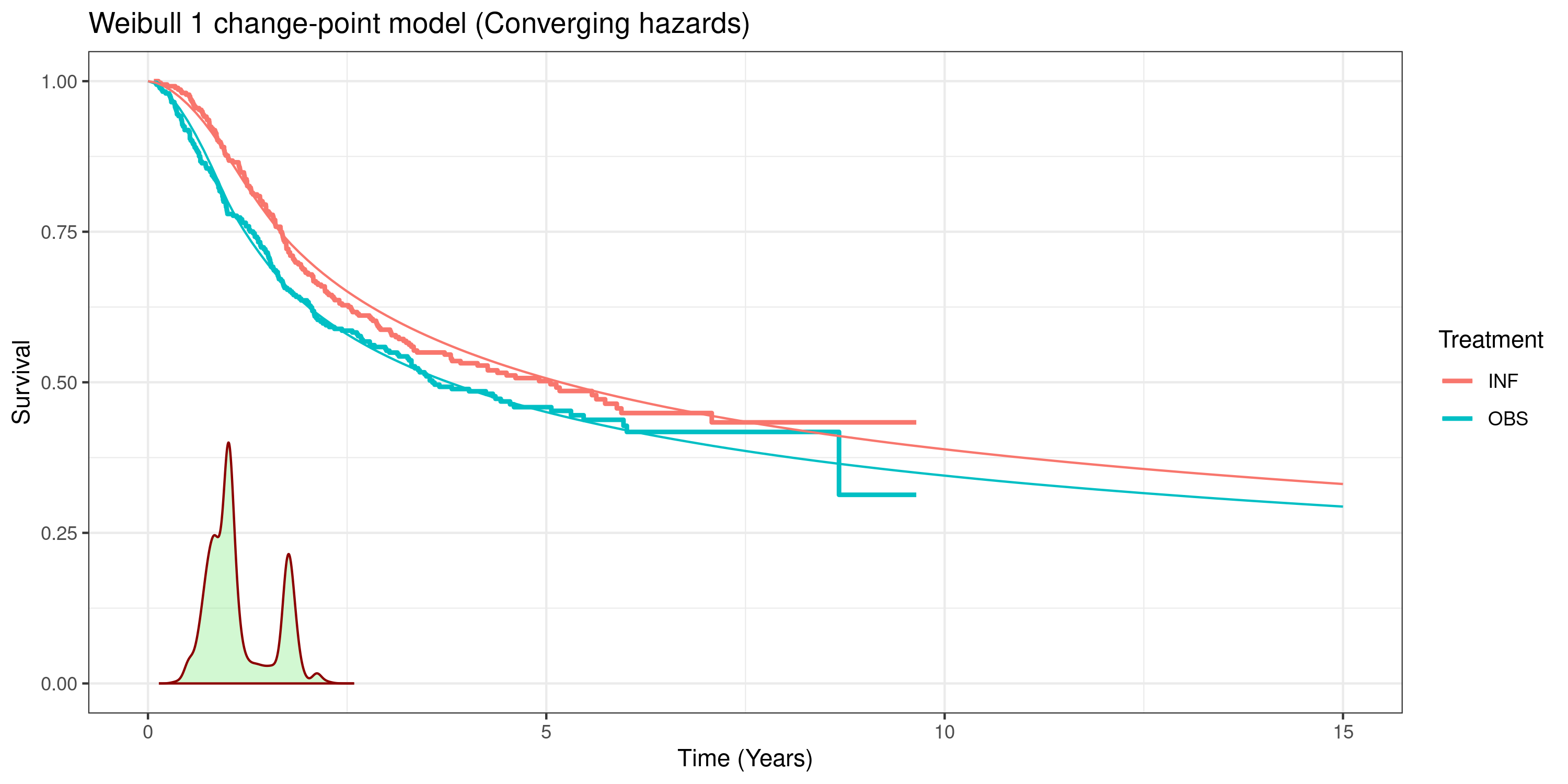}
            \caption{Scenario 3}%
            \label{fig:Scenario3}
        \end{subfigure}
        \hfill
        \begin{subfigure}[b]{0.475\textwidth}
            \centering
            \includegraphics[width=\textwidth]{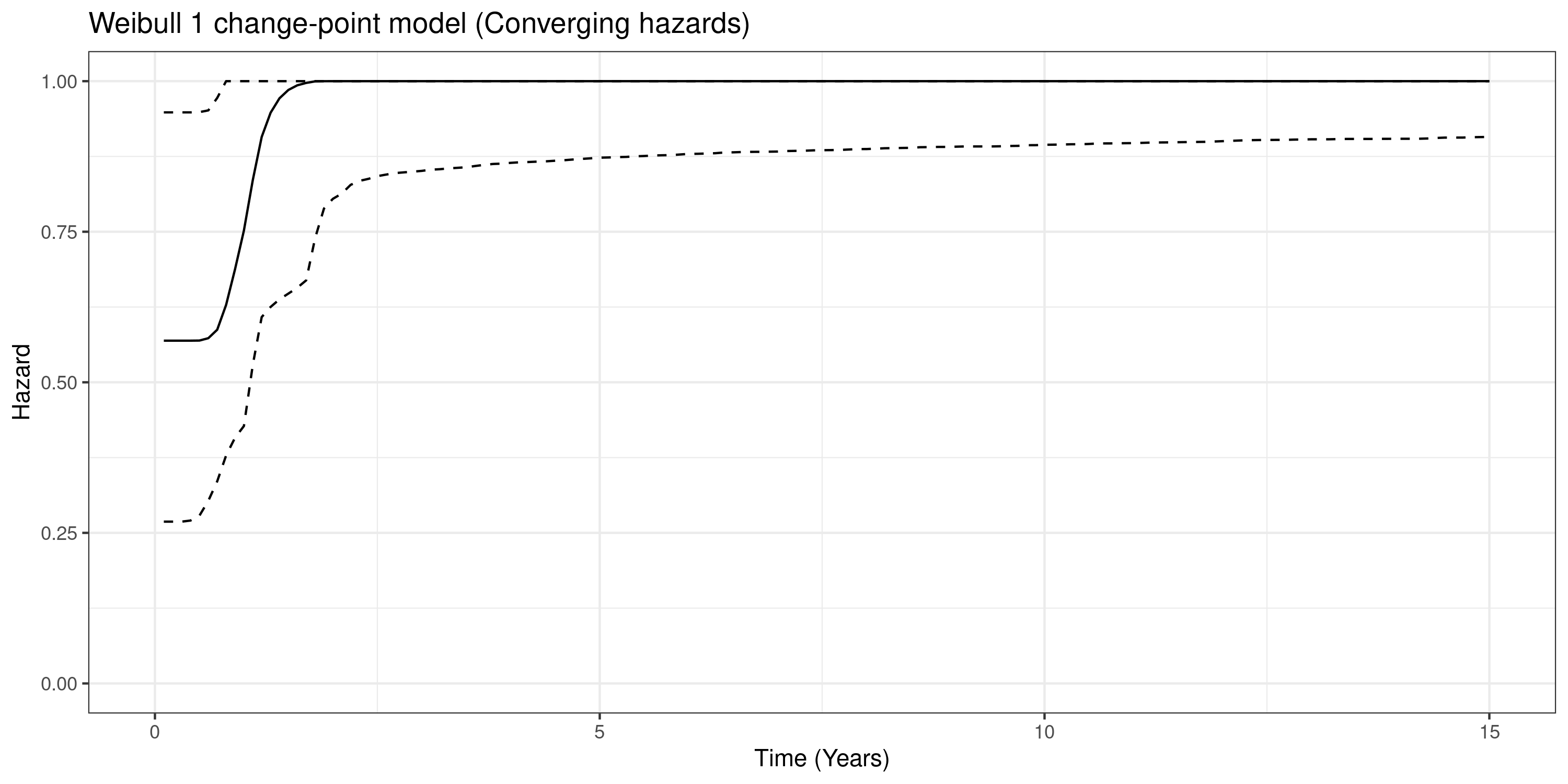}
            \caption{Scenario 3 - HR}%
            \label{fig:Scenario3-HR}
        \end{subfigure}
        \caption{E1690 \& E1684 - Various Change-point Hypotheses}
        \label{fig:E1690-E1684-All-Scenarios}
    \end{figure}

Figure \ref{fig:Scenario3-HR} shows the posterior distribution of the HR over time and highlights that the HR converges quite rapidly with the median value of the HR converging to 1 before year 2.

\subsubsection*{Interpretation of results Scenarios 1-3 }

It is worth noting that each of the methods considered here provide quite similar extrapolated survival estimates, however, the WAIC values are different for the scenario with a change-point for the hazard ratio. The model with the lowest WAIC is the scenario with the converging hazards (Scenario 3 WAIC 2009.18), however, this effectively equal to the common hazard (Scenario 2; WAIC 2009.35), and then the scenario with a change-point for the hazard ratio (Scenario 1; WAIC 2011.09).  The $\text{RMST}_{\text{diff}}$ for Scenarios 2 and 3 are very similar. This is because both models constrain the HR for the extrapolated region to be 1 or very close to 1. It is interesting to note that the model which allowed the HR to be unconstrained for the second interval produced a median posterior HR $>$ 1. This results in the $\text{RMST}_{\text{diff}}$ being lower for this scenario, however, the WAIC indicates that this additional parameter does not produce an improved goodness of fit relative to Scenario 2 (HR constrained to be = 1).

It is worth comparing the change-point survival models with flexible models which allow for non-proportional hazards. One such flexible parametric model is the Royston-Parmar cubic spline model which has the option to include ``knots'' to allow for flexible modelling of the baseline hazard function and which can estimate time-varying hazard ratios by allowing covariates (i.e. treatment status) on the higher-order terms (see $\texttt{flexsurvspline}$ for details).\cite{flexsurv.2016}  Although the flexible spline model (accommodating non-proportional hazards) visually fits the observed data quite well and has the lowest WAIC at 2003, the survival for the extrapolated region is unlikely to be plausible as there is crossing of the comparator arm (which was observation) with the actively treated arm (at 30 years the survival of the control arm vs the treatment arm is 27\% vs 19\%). It is worth noting that the standard Royston-Parmar PH model has a WAIC lower than the change-point models and has a similar $\text{RMST}_{\text{diff}}$ (See Table \ref{tab:RMST-E1690}).

\begin{table}[ht]
\centering
\caption{$\text{RMST}_{\text{diff}}$ for E1690 and E1684 datasets; All Scenarios}
\begin{tabular}{lrr}
  \hline
Model & $\text{RMST}_{\text{diff}}$ & WAIC \\
  \hline
  Royston Parmar (non-PH) & 0.23 & 2003.03 \\
  Royston Parmar (PH) & 0.78 & 2006.70 \\
  Change-point: Converging Hazards & 0.75 & 2009.18 \\
  Change-point: Equal Final Hazards & 0.67 & 2009.35 \\
  Change-point: HR (step) & 0.46 & 2011.09 \\
  Generalized-Gamma & 1.08 & 2012.50 \\
  Log-Normal & 0.93 & 2036.70 \\
  Log-Logistic & 0.79 & 2059.52 \\
  Gompertz & 0.75 & 2061.88 \\
  Weibull & 0.74 & 2098.73 \\
  Exponential & 0.74 & 2100.23 \\
  Gamma & 0.74 & 2101.65 \\
   \hline
\end{tabular}

\label{tab:RMST-E1690}
\end{table}

\subsection*{LUME-LUNG 1 - Delay of treatment effect}

\begin{figure}[H]
\centering
\begin{subfigure}[b]{0.5\textwidth}
  \centering
  \includegraphics[width=\textwidth]{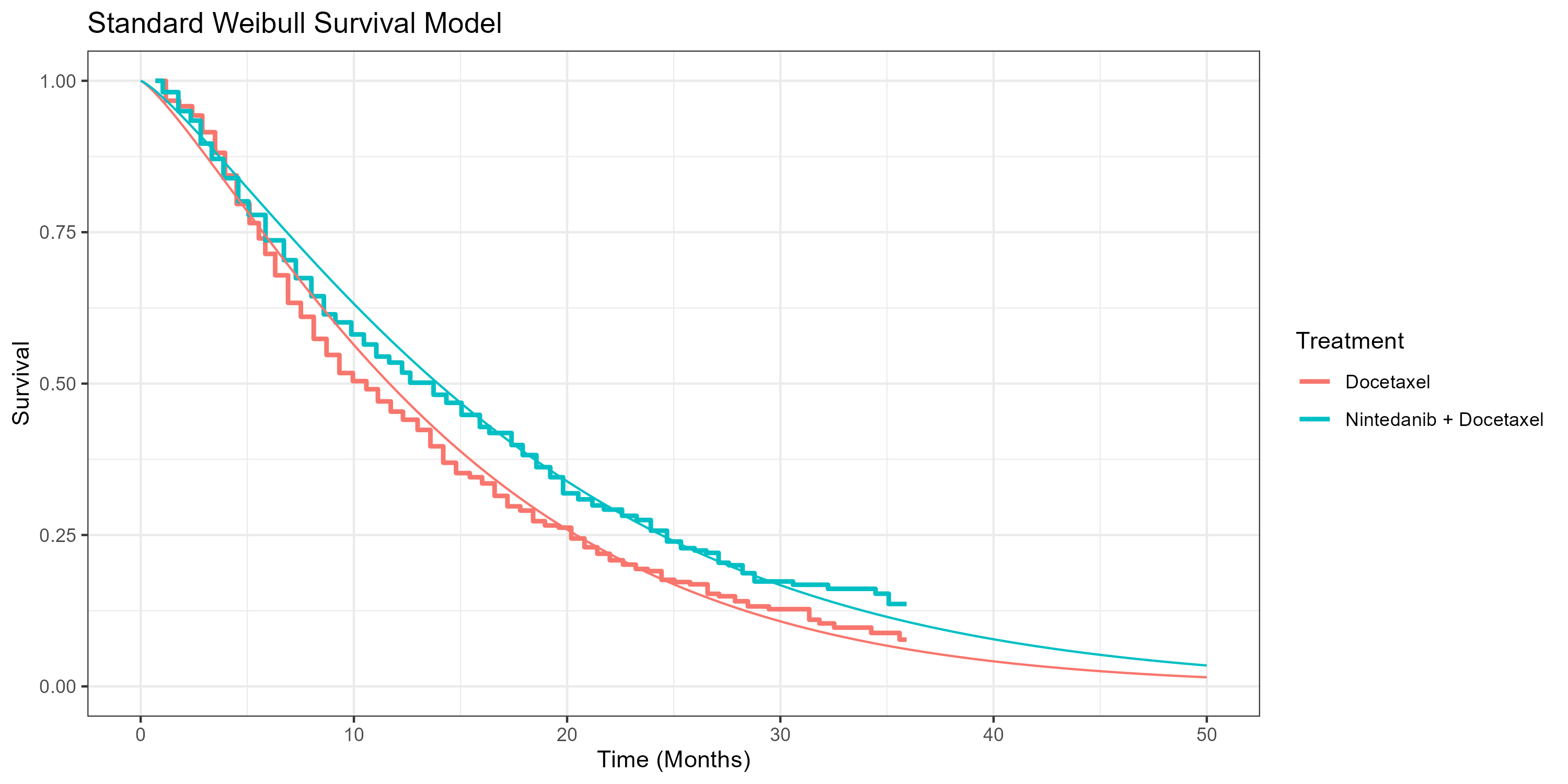}
  \caption{Standard Weibull Model}
  \label{fig:TA347-No-CP}
\end{subfigure}%
\begin{subfigure}[b]{0.5\textwidth}
  \centering
  \includegraphics[width=\textwidth]{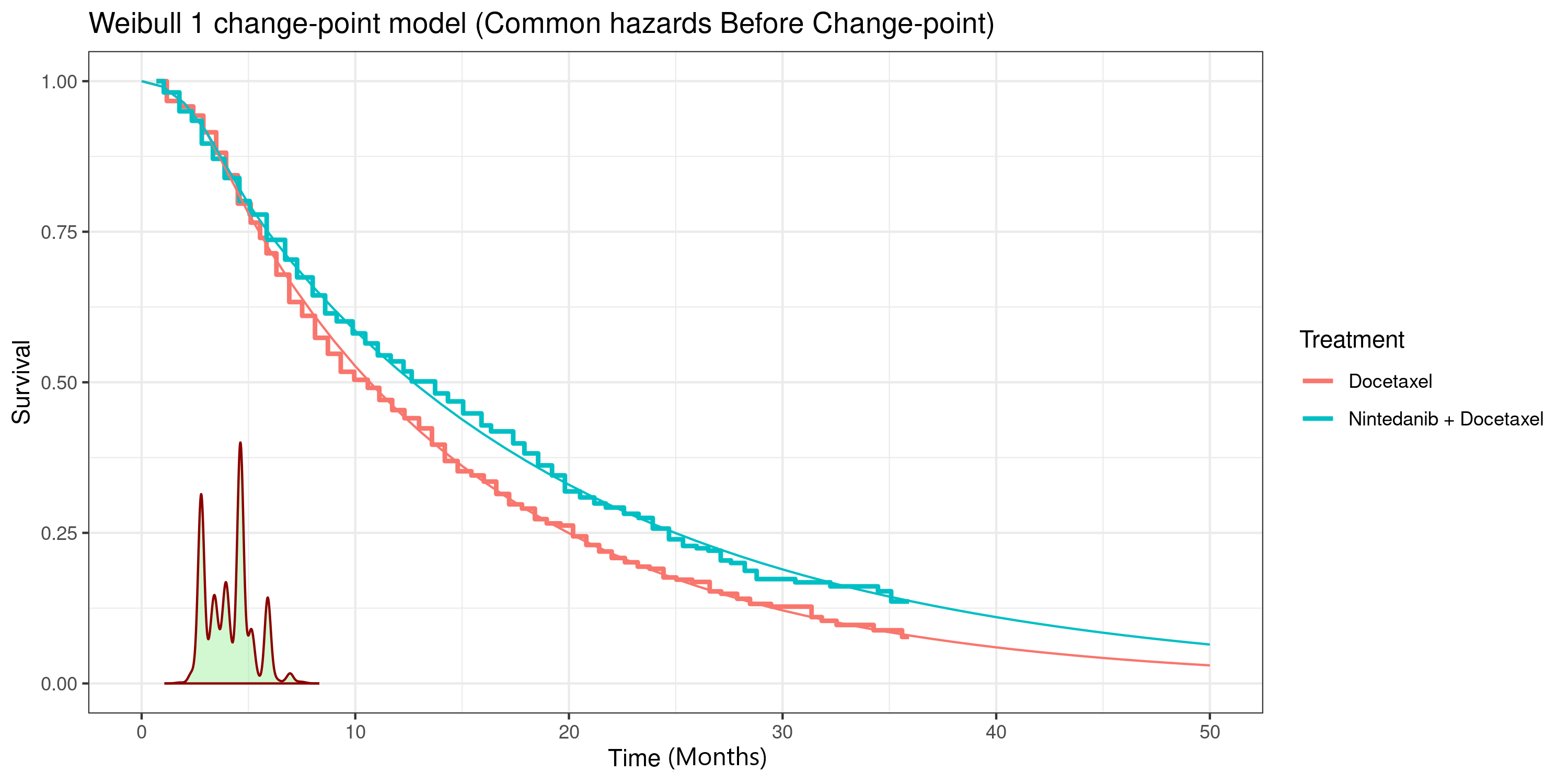}
  \caption{Changepoint Model}
  \label{fig:TA347-CP}
\end{subfigure}
\caption{LUME LUNG-1: No change-point and one change-point Weibull models}
\label{fig:fig:TA347}
\end{figure}

A Weibull model with a change-point assuming a common hazard before the change-point and a separate Weibull model with assuming a proportional hazard model is presented in Figure \ref{fig:TA347-No-CP}.

The standard Weibull model appears not to fit the data very well, particularly for the earlier part of the data. Allowing for the a change-point before which the hazards are equal and after which we assume proportional hazards appears to be a better fit to the data (Figure \ref{fig:TA347-CP}). In both the standard and change-point model the HR is $\approx 0.8$ between the treatment and comparator. The WAIC for the standard parametric model was 3970 while the WAIC for the change-point model was lower at 3933 (Table \ref{tab:RMST-LUME}).

\begin{table}[ht]
\centering
\caption{$\text{RMST}_{\text{diff}}$ for LUME-LUNG Treatment delay}
\begin{tabular}{lrr}
  \hline
Model & $\text{RMST}_{\text{diff}}$ & WAIC \\
  \hline
  Change-point & 2.72 & 3933.05 \\
  Log-Normal & 1.83 & 3935.17 \\
  Generalized-Gamma & 1.95 & 3936.63 \\
  Royston Parmar (PH) & 2.63 & 3937.92 \\
  Royston Parmar (non-PH) & 2.69 & 3939.61 \\
  Log-Logistic & 2.09 & 3945.81 \\
  Gamma & 2.56 & 3958.73 \\
  Weibull & 2.60 & 3969.97 \\
  Gompertz & 2.53 & 3995.38 \\
  Exponential & 2.51 & 4001.31 \\
   \hline
\end{tabular}
\label{tab:RMST-LUME}
\end{table}

\subsection*{BRIM-3 Study - Loss of Treatment effect}

We assume a change-point model in which a change-point is considered for the vermurafenib arm and after the change-point the hazard function is equal between vermurafenib and dacarbazine. We consider two change-point models, one assuming constant hazards for each segment compared with a model assuming a Weibull model for each segment. 



\begin{figure}[H]
\centering
\begin{subfigure}[b]{0.5\textwidth}
  \centering
  \includegraphics[width=\textwidth]{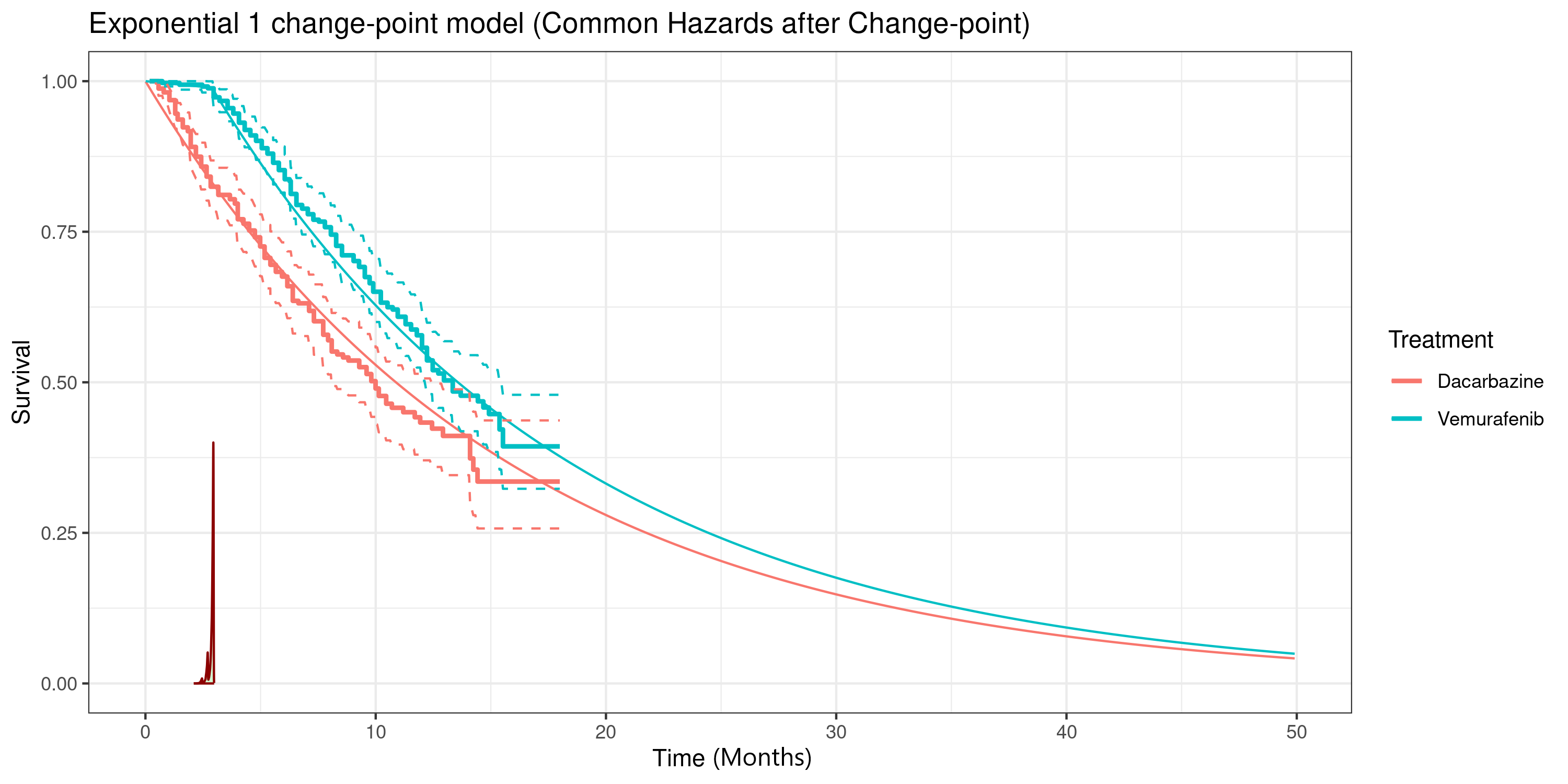}
  \caption{Change-point model with constant hazards.}
  \label{fig:BRIM3-Expo}
\end{subfigure}%
\begin{subfigure}[b]{0.5\textwidth}
  \centering
  \includegraphics[width=\textwidth]{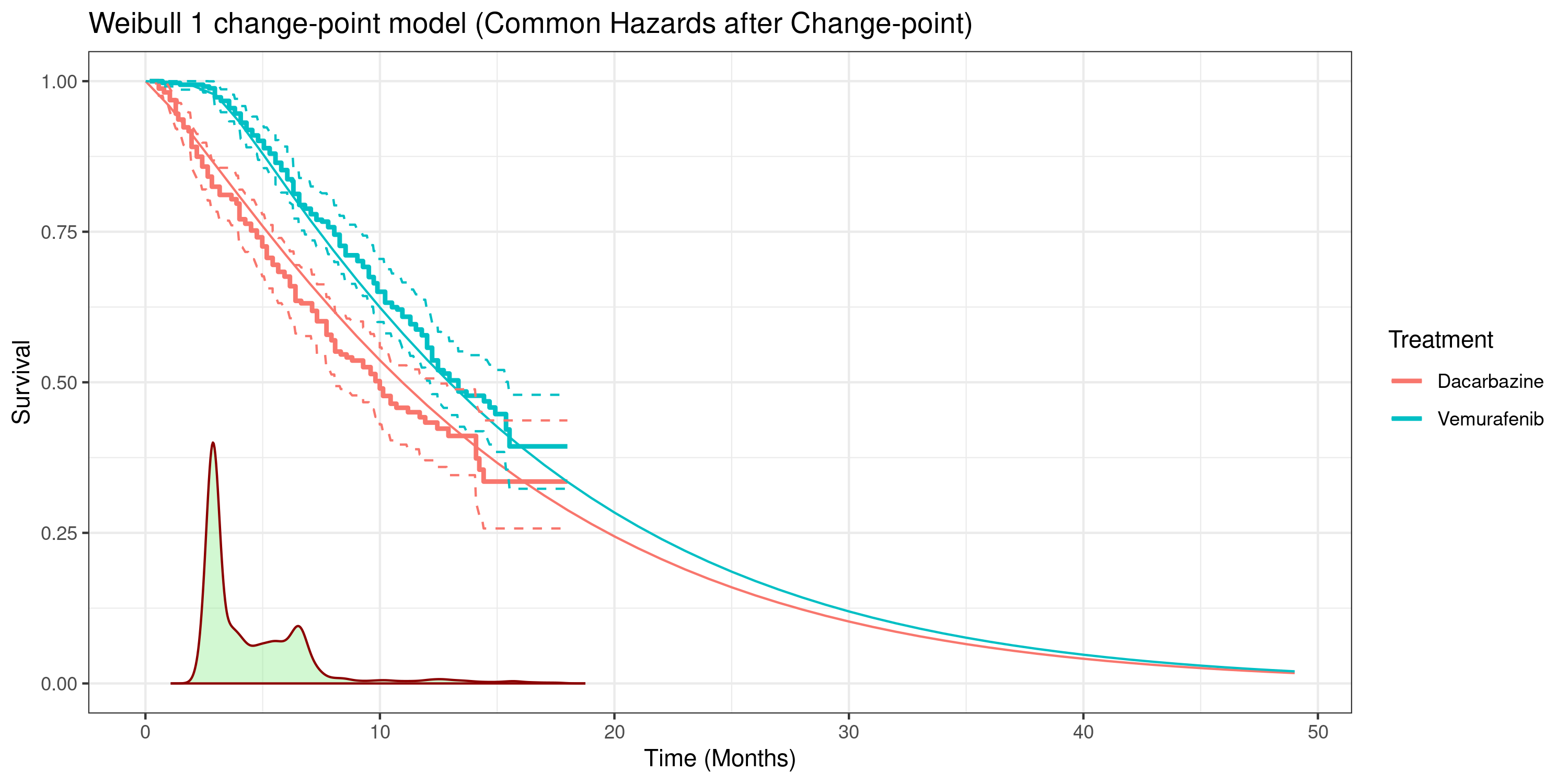}
  \caption{Change-point Model}
  \label{fig:BRIM3-Weibull}
\end{subfigure}
\caption{Changepoint scenarios for BRIM-3 survival data}
\label{fig:TA269}
\end{figure}

Comparing this model to the model assuming a hazards generated from a Weibull distribution we see that the survival is very similar between this model and the exponential survival model for the observed portion of the data, however, the extrapolated survival is different (Figures \ref{fig:BRIM3-Expo}, \ref{fig:BRIM3-Weibull}). This is because the Weibull model assumes a common monotonically increasing (as opposed to a constant) hazard resulting in a more rapid convergence in the survival curves\footnote{In theory the survival curves will only be equal in the limit as $t \to \infty$ so that $S(\infty)=0$, however, they are practically indistinguishable at 50 months}.

Although the models do not exactly match the Kaplan-Meier estimator for the later timepoints it should be noted that the expected survival function remains within the 95\% confidence intervals for the Kaplan-Meier trial and have a much lower WAIC than the other parametric models. The WAIC value for both models are very similar and slightly lower for the exponential change-point model, supporting the assertion of constant common hazards.\cite{Bagust.2014} It is worth noting that the posterior distribution for the change-point in the exponential model is very concentrated around 4 months, while for the more flexible Weibull model is much more diffuse. The best fitting model to the data is the Royston Parmar model with non-proportional hazards which fits the observed data well, however, the survival curves quickly cross and the negative $\text{RMST}_{\text{diff}}$ indicates that the over the time horizon (50 months) the expected survival is larger for the dacarbazine arm which is clearly implausible.

\begin{table}[H]
\centering
\caption{$\text{RMST}_{\text{diff}}$ for BRIM-3 Loss of Treatment effect}
\begin{tabular}{lrr}
  \hline
Model & $\text{RMST}_{\text{diff}}$ & WAIC\\
  \hline
  Royston Parmar (non-PH) & -0.51 & 2300.18 \\
  Change-point Exponential & 2.56 & 2306.92 \\
  Change-point Weibull & 1.94 & 2308.60 \\
  Log-Normal & 6.73 & 2312.84 \\
  Generalized-Gamma & 6.84 & 2314.77 \\
  Log-Logistic & 5.58 & 2317.90 \\
  Gamma & 5.10 & 2323.74 \\
  Royston Parmar (PH) & 5.04 & 2328.00 \\
  Weibull & 4.54 & 2330.37 \\
  Gompertz & 3.38 & 2351.04 \\
  Exponential & 5.23 & 2379.60 \\
   \hline
\end{tabular}
\label{tab:RMST-BRIM}
\end{table}

\newpage
\section*{Discussion}

In this chapter we have described a general class of survival change-point models and their estimation using modern Bayesian statistical software. Change-point models are particularly useful when modelling data with complex survival functions and when jointly modelling the intervention and comparator in instances when proportional hazards assumption fails. Through simulation studies we have seen that the change-point models produce most accurate extrapolations when the HR between treatment and comparator is substantial along with a large sample size. This is unsurprising as more complex data generating processes require a large number of observations to accurately estimate their underlying parameters. Because change-point models have comparatively more parameters, the associated likelihood surface estimated by these models can be relatively flat, particularly for datasets with high degree of censoring. This is particularly true for the converging hazard model.

Considering real examples, we analysed a large clinical trial dataset with a long follow-up and covariates (other than treatment status). We considered a variety of scenarios to jointly the model the treatment arms, finding that the relative treatment effect decreased over time. The various approaches to modelling survival produced similar and plausible extrapolated survival, in contrast to the non-proportional hazard flexible spline models which optimized the fit to the observed data but failed to produce sensible extrapolations i.e. crossing survival functions.


In the LUME-LUNG 1 dataset it appeared from the plot of the empirical survival that the survival did not diverge for a period of time after baseline, however, it is important to assess if the change-point model improved model fit to justify the inclusion of the four additional parameters (baseline shape and scale, change-point and HR for second interval). The final example models the hypothesis that a change-point model is present in one arm followed by common hazard applied to both arms. Previously when a common hazard model was suggested, the authors justified it by visual investigation of the cumulative hazard plots.\cite{Bagust.2014} By considering a change-point model we fully propagate statistical uncertainty while also testing an alternative hypothesis that a Weibull change-point model could have generate the data. We see that assuming Weibull model does have an impact on the extrapolated survival, however, based on goodness of fit it appears the exponential change-point model was most appropriate for the data.

Change-point models developed in this paper provide a consistent approach to the application of treatment effect waning assumptions which are often a source of disagreement between the company and decision makers in health technology assessments.~\cite{Kamgar.2022} Furthermore, the fact that uncertainty in the change-point was not fully propagated was a key concern raised by the decision maker.\cite{TA589}

The key advantage of change-point models is the flexibility to model a wide variety of scenarios, however, this flexibility does increase the number of potential models . Owing to the presence of potentially many competing scenarios, the modelled hypothesis selected as the basecase should undergo some clinical validation to assess the plausibility of the hypothesis. For example if a common hazard is to be assumed, an expert's opinion may be consulted to assess the timeframe within which this is most probable, and their beliefs formally integrated with the analysis. As mentioned previously change-point models are parameter rich and may be weakly identified from the data. Because of this it is useful to check the posterior distribution of the change-point to see if it has been reasonably informed by the inclusion of data. In contrast a relatively flat posterior distribution for the change-points  is suggestive of a weakly informed model.

One criticism of piecewise models by stating that only the final interval informs the extrapolation.\cite{rutherford.2020} In the case of change-point models, while this is also generally true, it does not have to be the case. For example with the converging hazard model the extrapolation is clearly informed by the hazard ratio from the first interval. Additionally we can specify the baseline hazard to be estimated from the entire time interval and only require the hazard ratios for the treatment arm to be interval specific. Even for the hazard ratio we could assume a hierarchical model whereby the prior for the hazard ratio for the current interval could be centered on the current value of the hazard ratio for the previous interval (as was considered in a piecewise exponential model with covariates).\cite{Ibrahim.2001}

Another criticism of change-point or piecewise models is that step changes in the hazard function are considered to be an implausible representation of the disease process.\cite{rutherford.2020} As was demonstrated in this paper, change-point models need not introduce discontinuities in the hazard function, however, in many situations such models produce a better model fit. As the hazard function is not an empirically observable quantity such as a survival function, therefore, we don't believe there is a strict requirement for continuity with the function. From our experience with real world survival data, there are many situations in which we observed the empirical survival function dropping precipitously and to attempt to model this with a continuous function may be unrealistic.
Change-point models could be  a useful approach to modelling a variety of hypothesis regarding relative treatment effectiveness. They enable the survival function to be accurately modelled while still allowing enforcing plausible extrapolations for the treatment arms. Although computationally more burdensome than standard parametric models (although by discretizing the time horizon these model can be estimated very quickly) we have shown how these models can be estimated using standard Bayesian software and provide fully worked examples for practitioners to apply to their own datasets. Further research will focus on estimation strategies which improve computational efficiency of these methods and developing a fully functioning R package similar to the $\texttt{mcp}$ package.\cite{mcp.2020}


\newpage
\bibliographystyle{ieeetr} 
\bibliography{References}

\appendix
\newpage

\section*{Pseudo Code for Change-point Models}
\subsection*{Parametric Change-point Model}

\begin{listing}[H]
\begin{minted}[fontsize=\footnotesize,breaklines,framerule=1pt]{C}
model {
  # Assuming a one change-point Weibull model with no covariates
  # Uniform prior for shape and scale.
  # Constant for zerors trick
  C <- 10000
  N_CP <- 1
  cp[1] = 0  # Should be zero
  cp[2] ~ dunif(0, max(time)) # Alternatives possible
  cp[3] = 100  # mcp helper value. very large number
  #Prior for the model parameters
  for(k in 1:(N_CP+1)){
      shape[k] ~ dunif(0,10)
      scale[k] ~ dunif(0,10)
  }
  # Model and likelihood
  for (i in 1:N) {
    for(k in 1:(N_CP+1)){
    #variable which gives the difference between the two intervals if time[i]>cp[k+1]
    #(i.e. cp[k+1] -cp[k]) or time between time[i] and cp[k]
    X[i,k] = max(min(time[i], cp[k+1]) - cp[k],0)
    #Indicator variable which highlights which interval time is in
    X_ind[i,k] = step(time[i]-cp[k])*step(cp[k+1]-time[i])

    log_haz_seg[i,k] <-  log(shape[k]*scale[k]*pow(time[i],shape[i,k]-1))*X_ind[i,k]
    cum_haz_seg[i,k] <- scale[i,k]*pow(X[i,k]+cp[k],shape[i,k]) -
                        scale[i,k]*pow(cp[k],shape[i,k])
    }
    log_haz[i] <- sum(log_haz_seg[i,])
    cum_haz[i] <- sum(cum_haz_seg[i,])
    loglik[i] = status[i]*log_haz[i] - cum_haz[i]
    #Zero Trick
    zero[i] ~ dpois(C - loglik[i])
  }
}
\end{minted}
\caption{Pseudo-Code for JAGS Change-point Model}
\label{code:JAGS-cp-code}
\end{listing}

\section*{Simulation Study}
\subsection*{Survival and Hazard functions of change-point models investigated in the simulation studies}\label{appendix:Sim-Study}

\begin{figure}[H]
    \centering
    \includegraphics[width=\textwidth]{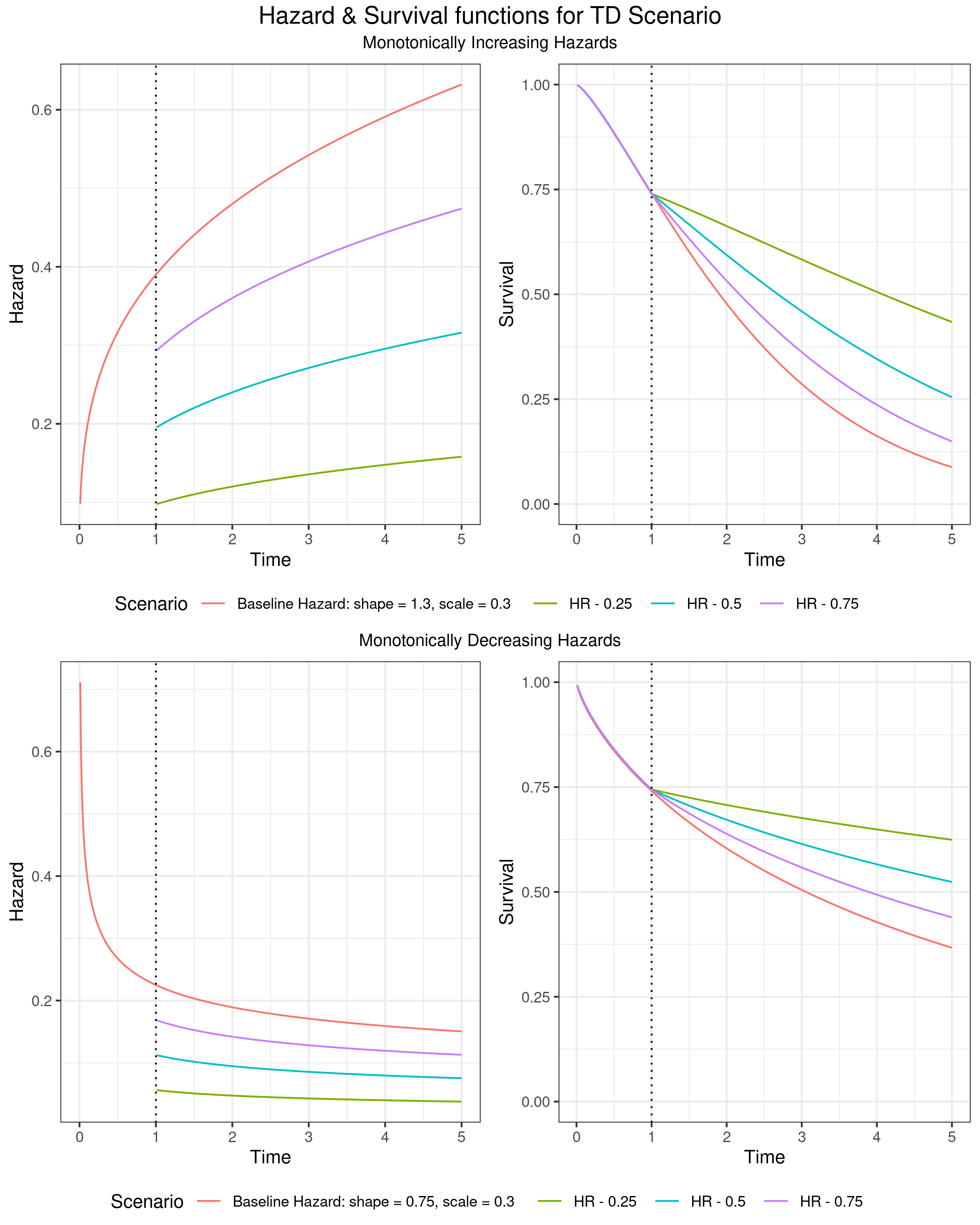}
    \caption{Hazard and Survival functions for Treatment Delay Scenarios}
    \label{fig:Scenario-TD}
\end{figure}

\begin{figure}[H]
    \centering
    \includegraphics[width=\textwidth]{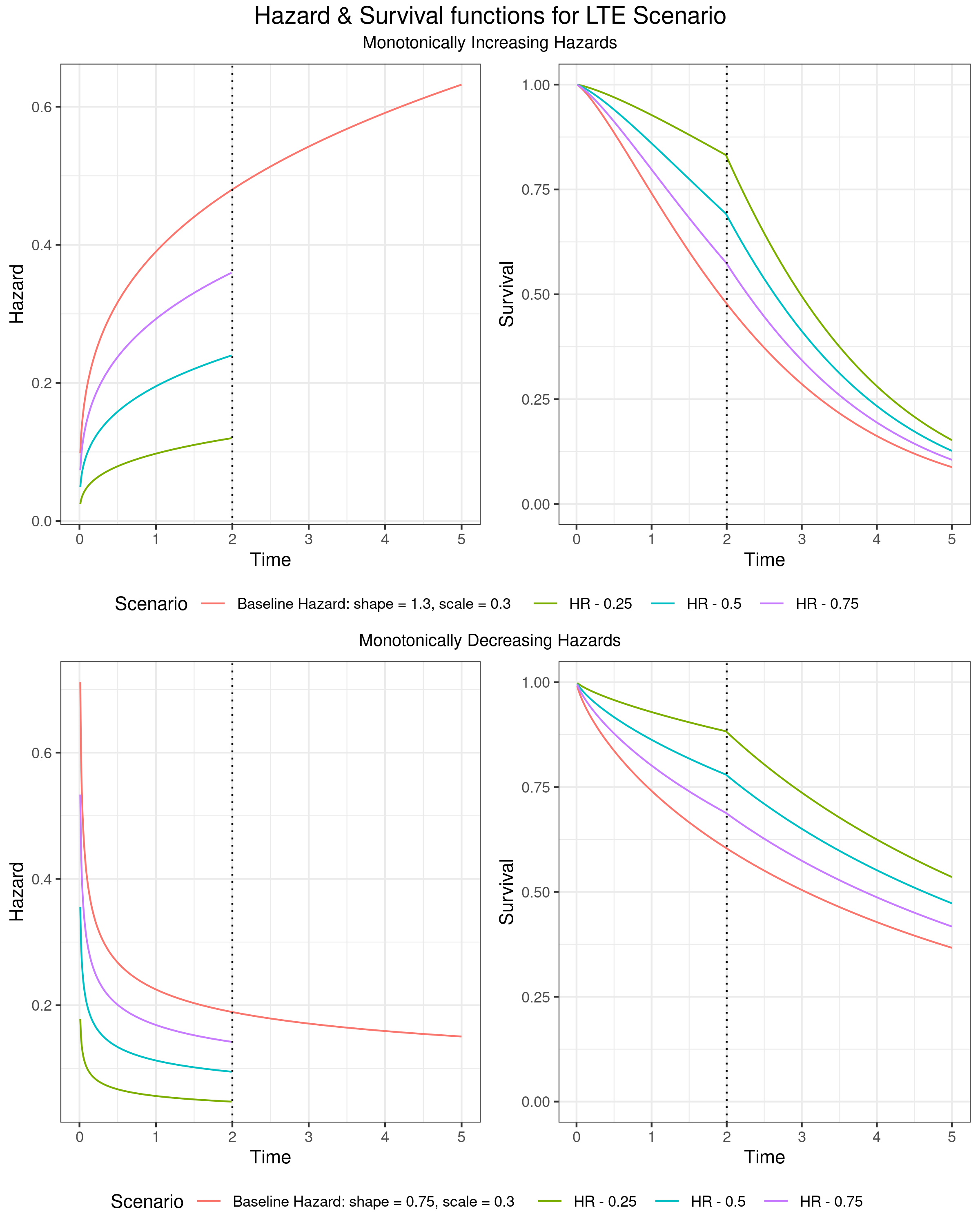}
    \caption{Hazard and Survival functions for Loss of Treatment Effect Scenarios}
    \label{fig:Scenario2-LTE}
\end{figure}

\begin{figure}[H]
    \centering
    \includegraphics[width=\textwidth]{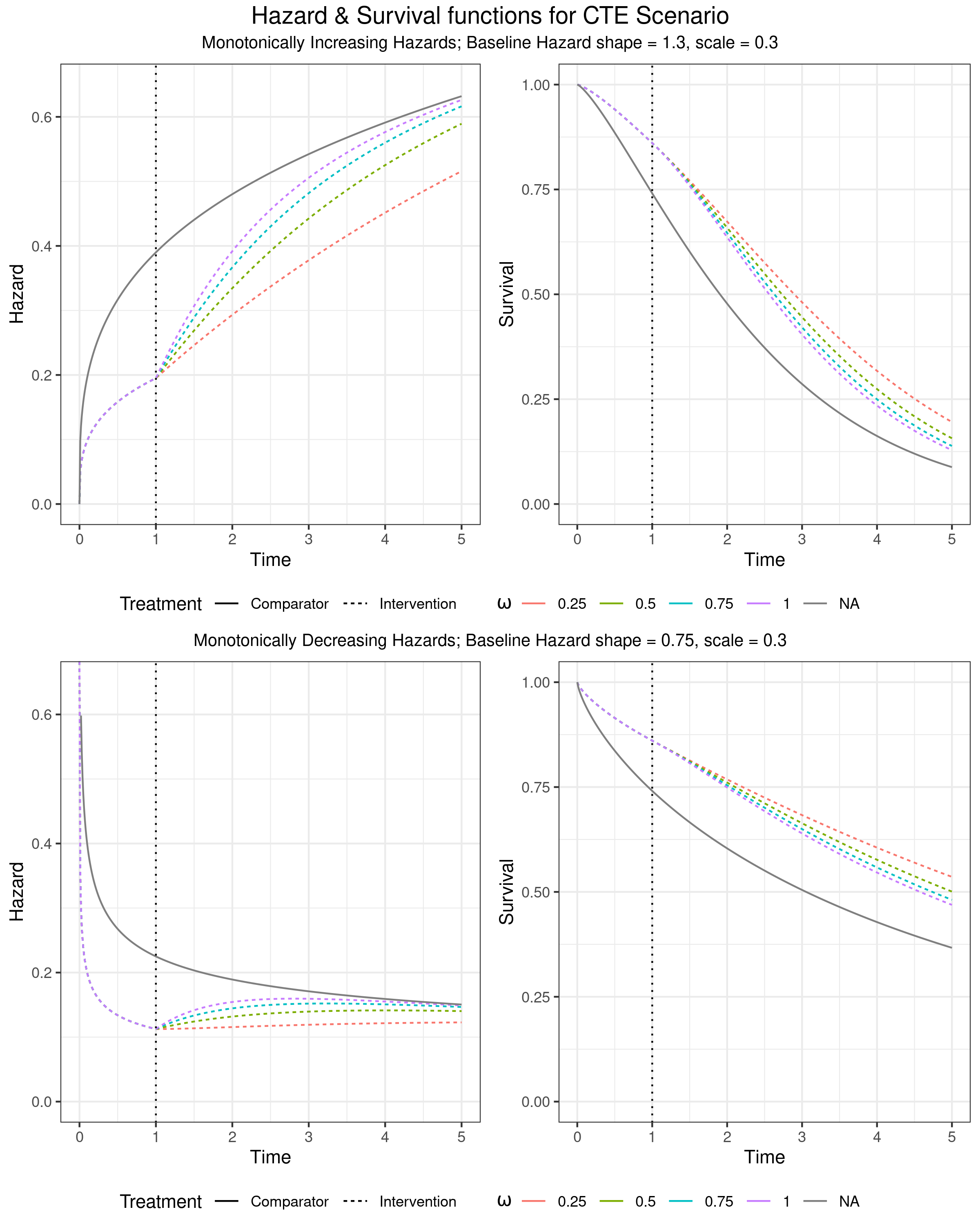}
    \caption{Hazard and Survival functions for Converging Treatment Effect Scenarios}
    \label{fig:Scenario3-CTE}
\end{figure}

\newpage
\subsection*{Simulation Study Results}\label{Appendix:Sim-Res}

\begin{table}[h!]
\centering
\caption{\label{CP-Scenario1} Simulation Study results - $\text{Err}_{\text{diff}}$  for Treatment Delay scenarios}
\begin{tabular}{|llll|lll|}
\hline
$n_\text{samp}$ & $t_\text{cens}$ & HR   & shape & Change-point model & Royston Parmar (non PH) & Gamma \\ \hline
100 & 3 & 0.25 & 1.3 & 0.23 & 0.41 & 0.26 \\
300 & 3 & 0.25 & 1.3 & 0.12 & 0.38 & 0.23 \\
500 & 3 & 0.25 & 1.3 & 0.08 & 0.36 & 0.23 \\
100 & 5 & 0.25 & 1.3 & 0.11 & 0.22 & 0.09 \\
300 & 5 & 0.25 & 1.3 & 0.06 & 0.21 & 0.06 \\
500 & 5 & 0.25 & 1.3 & 0.05 & 0.19 & 0.06 \\
100 & 3 & 0.5 & 1.3 & 0.32 & 0.43 & 0.25 \\
300 & 3 & 0.5 & 1.3 & 0.17 & 0.34 & 0.17 \\
500 & 3 & 0.5 & 1.3 & 0.1 & 0.3 & 0.15 \\
100 & 5 & 0.5 & 1.3 & 0.16 & 0.19 & 0.11 \\
300 & 5 & 0.5 & 1.3 & 0.08 & 0.15 & 0.07 \\
500 & 5 & 0.5 & 1.3 & 0.05 & 0.12 & 0.05 \\
100 & 3 & 0.75 & 1.3 & 0.52 & 0.62 & 0.4 \\
300 & 3 & 0.75 & 1.3 & 0.26 & 0.33 & 0.25 \\
500 & 3 & 0.75 & 1.3 & 0.2 & 0.29 & 0.2 \\
100 & 5 & 0.75 & 1.3 & 0.28 & 0.22 & 0.18 \\
300 & 5 & 0.75 & 1.3 & 0.15 & 0.14 & 0.12 \\
500 & 5 & 0.75 & 1.3 & 0.14 & 0.15 & 0.13 \\
100 & 3 & 0.25 & 0.75 & 0.22 & 0.19 & 0.34 \\
300 & 3 & 0.25 & 0.75 & 0.1 & 0.13 & 0.33 \\
500 & 3 & 0.25 & 0.75 & 0.08 & 0.13 & 0.34 \\
100 & 5 & 0.25 & 0.75 & 0.14 & 0.1 & 0.13 \\
300 & 5 & 0.25 & 0.75 & 0.06 & 0.06 & 0.12 \\
500 & 5 & 0.25 & 0.75 & 0.06 & 0.05 & 0.13 \\
100 & 3 & 0.5 & 0.75 & 0.1 & 0.21 & 0.38 \\
300 & 3 & 0.5 & 0.75 & 0.18 & 0.19 & 0.3 \\
500 & 3 & 0.5 & 0.75 & 0.14 & 0.15 & 0.32 \\
100 & 5 & 0.5 & 0.75 & 0.2 & 0.18 & 0.17 \\
300 & 5 & 0.5 & 0.75 & 0.12 & 0.1 & 0.14 \\
500 & 5 & 0.5 & 0.75 & 0.08 & 0.09 & 0.12 \\
100 & 3 & 0.75 & 0.75 & 0.82 & 0.66 & 0.56 \\
300 & 3 & 0.75 & 0.75 & 0.47 & 0.46 & 0.41 \\
500 & 3 & 0.75 & 0.75 & 0.33 & 0.32 & 0.31 \\
100 & 5 & 0.75 & 0.75 & 0.62 & 0.44 & 0.39 \\
300 & 5 & 0.75 & 0.75 & 0.26 & 0.25 & 0.23 \\
500 & 5 & 0.75 & 0.75 & 0.22 & 0.18 & 0.2 \\

\hline
\end{tabular}
\end{table}

\begin{table}[h!]
\centering
\caption{\label{CP-Scenario2} Simulation Study results - $\text{Err}_{\text{diff}}$  for Loss of Treatment Effect scenarios}

\begin{tabular}{|llll|lll|}
\hline
$n_\text{samp}$ & $t_\text{cens}$ & HR   & shape & Change-point model & Royston Parmar (non-PH) & Royston Parmar (PH) \\ \hline
100 & 3 & 0.25 & 1.3 & 0.32 & 0.51 & 0.28 \\
300 & 3 & 0.25 & 1.3 & 0.16 & 0.41 & 0.16 \\
500 & 3 & 0.25 & 1.3 & 0.13 & 0.42 & 0.12 \\
100 & 5 & 0.25 & 1.3 & 0.16 & 0.17 & 0.19 \\
300 & 5 & 0.25 & 1.3 & 0.08 & 0.12 & 0.15 \\
500 & 5 & 0.25 & 1.3 & 0.06 & 0.11 & 0.16 \\
100 & 3 & 0.5 & 1.3 & 0.46 & 0.59 & 0.38 \\
300 & 3 & 0.5 & 1.3 & 0.23 & 0.38 & 0.21 \\
500 & 3 & 0.5 & 1.3 & 0.17 & 0.34 & 0.17 \\
100 & 5 & 0.5 & 1.3 & 0.26 & 0.25 & 0.26 \\
300 & 5 & 0.5 & 1.3 & 0.13 & 0.14 & 0.16 \\
500 & 5 & 0.5 & 1.3 & 0.09 & 0.11 & 0.14 \\
100 & 3 & 0.75 & 1.3 & 0.22 & 0.27 & 0.4 \\
300 & 3 & 0.75 & 1.3 & 0.54 & 0.68 & 0.49 \\
500 & 3 & 0.75 & 1.3 & 0.45 & 0.46 & 0.35 \\
100 & 5 & 0.75 & 1.3 & 0.59 & 0.45 & 0.45 \\
300 & 5 & 0.75 & 1.3 & 0.27 & 0.25 & 0.26 \\
500 & 5 & 0.75 & 1.3 & 0.27 & 0.25 & 0.26 \\
100 & 3 & 0.25 & 0.75 & 0.02 & 0.02 & 0.02 \\
300 & 3 & 0.25 & 0.75 & 0.33 & 0.89 & 0.74 \\
500 & 3 & 0.25 & 0.75 & 0.24 & 0.76 & 0.7 \\
100 & 5 & 0.25 & 0.75 & 0.35 & 0.55 & 0.36 \\
300 & 5 & 0.25 & 0.75 & 0.19 & 0.44 & 0.29 \\
500 & 5 & 0.25 & 0.75 & 0.15 & 0.42 & 0.3 \\
100 & 3 & 0.5 & 0.75 & 0.87 & 0.87 & 0.86 \\
300 & 3 & 0.5 & 0.75 & 0.51 & 0.63 & 0.76 \\
500 & 3 & 0.5 & 0.75 & 0.35 & 0.59 & 0.71 \\
100 & 5 & 0.5 & 0.75 & 0.56 & 0.59 & 0.48 \\
300 & 5 & 0.5 & 0.75 & 0.29 & 0.39 & 0.35 \\
500 & 5 & 0.5 & 0.75 & 0.22 & 0.27 & 0.35 \\
100 & 3 & 0.75 & 0.75 & 2.21 & 1.95 & 1.4 \\
300 & 3 & 0.75 & 0.75 & 1.11 & 0.94 & 0.89 \\
500 & 3 & 0.75 & 0.75 & 0.92 & 0.86 & 0.86 \\
100 & 5 & 0.75 & 0.75 & 1.14 & 1.07 & 0.82 \\
300 & 5 & 0.75 & 0.75 & 0.58 & 0.6 & 0.55 \\
500 & 5 & 0.75 & 0.75 & 0.48 & 0.45 & 0.44 \\
\hline
\end{tabular}
\end{table}

\begin{table}[h!]
\centering
\caption{\label{CP-Scenario3} Simulation Study results - $\text{Err}_{\text{diff}}$  for Converging Treatment Effect scenarios}

\begin{tabular}{|lllll|lll|}
\hline
$n_\text{samp}$ & $t_\text{cens}$ & HR   & shape & $\omega$ & Change-point model & Royston Parmar (non-PH) & Gompertz \\ \hline
100 & 3 & 0.25 & 1.3 & 1 & 0.39 & 0.44 & 0.5 \\
300 & 3 & 0.25 & 1.3 & 1 & 0.24 & 0.29 & 0.4 \\
500 & 3 & 0.25 & 1.3 & 1 & 0.19 & 0.22 & 0.35 \\
100 & 5 & 0.25 & 1.3 & 1 & 0.37 & 0.38 & 0.5 \\
300 & 5 & 0.25 & 1.3 & 1 & 0.23 & 0.24 & 0.43 \\
500 & 5 & 0.25 & 1.3 & 1 & 0.17 & 0.19 & 0.4 \\
100 & 3 & 0.5 & 1.3 & 1 & 0.4 & NA & 0.44 \\
300 & 3 & 0.5 & 1.3 & 1 & 0.23 & 0.25 & 0.28 \\
500 & 3 & 0.5 & 1.3 & 1 & 0.18 & 0.21 & 0.26 \\
100 & 5 & 0.5 & 1.3 & 1 & 0.36 & 0.36 & 0.39 \\
300 & 5 & 0.5 & 1.3 & 1 & 0.23 & 0.23 & 0.3 \\
500 & 5 & 0.5 & 1.3 & 1 & 0.18 & 0.18 & 0.28 \\
100 & 3 & 0.75 & 1.3 & 1 & 0.39 & 0.45 & 0.4 \\
300 & 3 & 0.75 & 1.3 & 1 & 0.23 & 0.25 & 0.24 \\
500 & 3 & 0.75 & 1.3 & 1 & 0.19 & 0.2 & 0.2 \\
100 & 5 & 0.75 & 1.3 & 1 & 0.39 & 0.38 & 0.41 \\
300 & 5 & 0.75 & 1.3 & 1 & 0.22 & 0.22 & 0.23 \\
500 & 5 & 0.75 & 1.3 & 1 & 0.16 & 0.16 & 0.18 \\
100 & 3 & 0.25 & 0.75 & 1 & 0.17 & 0.17 & 0.19 \\
300 & 3 & 0.25 & 0.75 & 1 & 0.57 & 0.88 & 1.98 \\
500 & 3 & 0.25 & 0.75 & 1 & 0.48 & 0.73 & 2.1 \\
100 & 5 & 0.25 & 0.75 & 1 & 0.88 & 1.12 & 1.43 \\
300 & 5 & 0.25 & 0.75 & 1 & 0.53 & 0.64 & 1.29 \\
500 & 5 & 0.25 & 0.75 & 1 & 0.4 & 0.51 & 1.32 \\
100 & 3 & 0.5 & 0.75 & 1 & 0.95 & 1.26 & 1.51 \\
300 & 3 & 0.5 & 0.75 & 1 & 0.56 & 0.69 & 1.29 \\
500 & 3 & 0.5 & 0.75 & 1 & 0.46 & 0.68 & 1.31 \\
100 & 5 & 0.5 & 0.75 & 1 & 0.98 & 1.03 & 1.2 \\
300 & 5 & 0.5 & 0.75 & 1 & 0.46 & 0.53 & 0.86 \\
500 & 5 & 0.5 & 0.75 & 1 & 0.41 & 0.45 & 0.85 \\
100 & 3 & 0.75 & 0.75 & 1 & 0.92 & 1.06 & 1.1 \\
300 & 3 & 0.75 & 0.75 & 1 & 0.52 & 0.63 & 0.73 \\
500 & 3 & 0.75 & 0.75 & 1 & 0.44 & 0.56 & 0.74 \\
100 & 5 & 0.75 & 0.75 & 1 & 0.9 & 0.93 & 0.98 \\
300 & 5 & 0.75 & 0.75 & 1 & 0.47 & 0.5 & 0.59 \\
500 & 5 & 0.75 & 0.75 & 1 & 0.41 & 0.43 & 0.55 \\
\hline
\end{tabular}
\end{table}

\end{document}